\documentclass[nofootinbib,aps,prd,reprint,superscriptaddress,showkeys]{revtex4-2}

\usepackage[utf8]{inputenc} 
\usepackage{graphicx,overpic,mathtools}
\usepackage{amsthm,amsmath,amssymb,hyperref}
\usepackage{braket,bm,bbm,setspace}
\usepackage[normalem]{ulem} 
\usepackage{physics}
\usepackage{float}
\usepackage[makeroom]{cancel}
\usepackage[english]{babel}
\usepackage{xcolor}
\usepackage{tensor}
\graphicspath{ {images/} }
\addto\captionsspanish{}
\hypersetup{
	colorlinks=true,
	pdfborder={0 0 0},
	citecolor=purple,
	linkcolor=blue,
	filecolor=blue,
	urlcolor=blue,
}
\usepackage{subfigure}

\usepackage{tikz}
\usetikzlibrary{quantikz2}

\usepackage{booktabs}
\usepackage{multirow}

\usepackage{adjustbox}
\usepackage{enumerate}
\usepackage{orcidlink}

\makeatletter
\def\l@subsection#1#2{}
\def\l@subsubsection#1#2{}
\makeatother

\newcommand{\nT}{L}
\newcommand{\adder}{\mathcal{P}^+}
\newcommand{\update}{V}
\newcommand{\adderqubits}{m}

\begin{document}

\title{Implementing Semiclassical Szegedy Walks in Classical-Quantum Circuits for Homomorphic Encryption}
\author{Sergio A. Ortega \orcidlink{0000-0002-8237-7711}}
\email{sergioan@ucm.es}
\affiliation{Departamento de Física Teórica, Universidad Complutense de Madrid, 28040 Madrid, Spain}
\author{Pablo Fernández \orcidlink{0009-0008-5113-3248}}
\email{pabfer23@ucm.es}
\affiliation{Departamento de Física Teórica, Universidad Complutense de Madrid, 28040 Madrid, Spain}
\author{Miguel A. Martin-Delgado \orcidlink{0000-0003-2746-5062}}
\email{mardel@ucm.es}
\affiliation{Departamento de Física Teórica, Universidad Complutense de Madrid, 28040 Madrid, Spain}
\affiliation{CCS-Center for Computational Simulation, Universidad Politécnica de Madrid, 28660 Boadilla del Monte, Madrid, Spain.}

\begin{abstract}
As cloud services continue to expand, the security of private data stored and processed in these environments has become paramount. This work delves into quantum homomorphic encryption (QHE), an emerging technology that facilitates secure computation on encrypted quantum data without revealing the underlying information. We reinterpret QHE schemes through classical-quantum circuits, enhancing efficiency and addressing previous limitations related to key computations. Our approach eliminates the need for exponential key preparation by calculating keys in real-time during simulation, leading to a linear complexity in classically controlled gates. We also investigate the $T/T^{\dagger}$-gate complexity associated with various quantum walks, particularly Szegedy quantum and semiclassical algorithms, demonstrating efficient homomorphic implementations across different graph structures. Our simulations, conducted in Qiskit, validate the effectiveness of QHE for both standard and semiclassical walks. The rules for the homomorphic evaluation of the reset and intermediate measurement operations have also been included to perform the QHE of semiclassical walks. Additionally, we introduce the CQC-QHE library, a comprehensive tool that simplifies the construction and simulation of classical-quantum circuits tailored for quantum homomorphic encryption. Future work will focus on optimizing classical functions within this framework and exploring broader graph types to enhance QHE applications in practical scenarios.
\end{abstract}

\maketitle

\onecolumngrid

\section{Introduction}

Currently, plenty of users have access to the services offered by the cloud for many different purposes, such as storage. This means that enormous quantities of private data have been uploaded to the cloud. Thus, preserving the data's security is an extremely important task. In order to accomplish this, cryptographic technology should be used to protect any computations performed on the cloud. In this regard, a promising technology that has been developed in recent years is homomorphic encryption. It allows a client to encrypt data and then send it to a remote server that performs operations on this encrypted data. Once these operations are finished, the server returns the data back for decryption. This way the client receives the evaluated data while the server does not obtain any information about the actual data it operated with. The first classical fully homomorphic encryption scheme (FHE) was proposed by Gentry \cite{Gentry}. Since then, more classical schemes have been proposed and perfected. These schemes are computationally secure, which means that their security is based on the difficulty of solving mathematical problems, like the learning with errors problem (LWE) \cite{LWE} or ideal lattices \cite{Gentry}.\\

Quantum computers accessible through the cloud have already been developed, like the IBM quantum computers \cite{IBMQ}. Since most quantum computers will be accessible through the cloud in the near future, different quantum homomorphic encryption (QHE) schemes have been developed to preserve the computer's security \cite{Rhode,QFHE_def,Tan}. QHE allows a remote server to apply a quantum circuit on the encrypted quantum data that a client has provided. Then, the client decrypts the server's output and obtains the result of the quantum circuit without revealing the results of the computation to the server.\\

QHE schemes can be classified as either interactive or non-interactive. The scheme is said to be interactive if the client and the server communicate during the execution of the circuit. Interactions decrease the efficiency of the scheme since the server must wait until the client finishes some operations before it can continue executing the scheme \cite{Liang}. We are more interested in studying non-interactive schemes. Some important definitions regarding the different properties of QHE schemes are given below. A QHE scheme is $\mathcal{F}$-homomorphic if it can evaluate any quantum circuit homomorphically, so that the circuit can contain any gate of the universal Clifford+$T$ gates set. A QHE scheme is said to be compact if the complexity of its decryption procedure does not depend on the evaluated circuit. For a scheme to be considered a quantum fully homomorphic encryption (QFHE) scheme, it must be $\mathcal{F}$-homomorphic and compact. A scheme is perfectly secure if its security does not depend on the difficulty of solving complex mathematical problems, so the ciphertext conveys no information about the contents of the plaintext.\\

Questions regarding the theoretical limits of QHE schemes have been explored. A no-go result which states that any QFHE with perfect security must produce exponential storage overhead has been proved \cite{No_go_result}. Later, an enhanced no-go result which states that constructing a QFHE scheme that is both information theoretically secure (ITS) and non-interactive is impossible was proved \cite{Enhanced_no_go}. Then, the best security a non-interactive QFHE scheme can achieve is computational security.\\

Due to these no-go results no QHE scheme can have non-interaction, compactness, $\mathcal{F}$-homomorphism and perfect security simultaneously. For this reason different schemes have been proposed with less strict conditions, such as allowing interactions \cite{T_interactions}, downgrading the security level from perfect security to computational security \cite{Broadbent} or making the scheme not $\mathcal{F}$-homomorphic \cite{Enhanced_no_go}.\\

Two non-interactive quantum homomorphic schemes were constructed combining quantum one-time pad (QOTP) and classical FHE \cite{Broadbent}. The first scheme they constructed is known as EPR (Einstein, Podolski and Rosen \cite{Einstein}). Its properties are $\mathcal{F}$-homomorphism, non-interaction, computational security and quasi-compactness. A scheme is said to be quasi-compact if the complexity of its decryption procedure scales sublinearly in the size of the evaluated circuit \cite{Broadbent}. EPR was proved to be $L^2$-quasi-compact, where $L$ is the number of $T/T^\dagger$ gates contained in the evaluated circuit, so the complexity of its decryption procedure scales with the square of the number of $T/T^\dagger$ gates contained in the evaluated circuit.\\

Liang \cite{Liang} constructed two perfectly secure, non-interactive and $\mathcal{F}$-homomorphic QHE schemes that are not compact but quasi-compact. Both Broadbent's and Liang's schemes make use of quantum measurements and Bell states. One of Liang's schemes called VGT, is $L$-quasi-compact, so it is superior in that aspect compared to EPR. Since these schemes are quasi-compact, they do not contradict the no go-result presented previously.\\

The main feature of Liang's schemes \cite{Liang} is that although they are quasi-compact, they allow the efficient homomorphic evaluation of any quantum circuit with low $T/T^\dagger$-gate complexity with perfect data security. For this reason, they are suitable for circuits with a polynomial number of $T/T^{\dagger}$ gates. On the other hand, the decryption procedure would be inefficient for circuits containing an exponential number of $T/T^\dagger$ gates. Liang's schemes have been used for different purposes. Gong et al. \cite{QHE_grover} used these schemes to implement a ciphertext retrieval scheme based on Grover's algorithm. They performed experiments using Qiskit and IBM's quantum computers as an example of the scheme. A quantum ciphertext dimension reduction scheme for homomorphic encrypted data based on these schemes was constructed in \cite{Gong2}. In \cite{Pablo_bernstein} Liang's schemes are applied to the recursive Bernstein-Vazirani algorithm, showing different cases in which the QHE schemes are efficient.\\

Regarding the simulation of QHE schemes, an open-source Python implementation of EPR has recently been provided \cite{QHE_simulation_EPR}. Circuits consisting only on Clifford gates were also simulated using Qiskit and Broadbent's schemes \cite{QHE_qiskit_clifford}. As we mentioned, Liang's schemes were also simulated in different works. A simulation of Grover's search was performed using a circuit that only needed 2 qubits and Clifford gates \cite{QHE_grover}. Moreover, they performed an experimental implementation of the decryption scheme for a circuit containing a single $T$ gate using an IBM quantum computer. A more complex Grover search circuit encrypted with Liang's schemes was simulated in Qiskit \cite{Pablo_Grover}. This simulation contained three qubits and seven $T/T^{\dagger}$ gates. Using classically controlled $S$ gates, they managed to correct the errors generated by the $T/T^{\dagger}$ gates. However, the value of each key has to be calculated beforehand to determine the correct control value of the classically controlled $S$ gates. This requires the calculation of a number of key values that grows exponentially with $L$, the number of $T/T^{\dagger}$ gates, and an exponential number of classically controlled $S$ gates.\\

In this paper, we reformulate Liang's schemes using classical-quantum circuits for the decryption procedure. This reinterpretation of the scheme allows us to construct a classical simulator using Qiskit. Compared to the previous simulations of Liang's schemes our simulator constitute an improvement, since it does not require any previous calculation of the keys. Instead, the keys are calculated at running time. This results in a simulation with a linear number of classically controlled $S$ gates instead of exponential so that the circuits can contain any arbitrary number of $T/T^{\dagger}$ gates. Based on our simulation algorithm, we have also developed a Python library named Classical-Quantum Circuits for Quantum Homomorphic Encryption (CQC-QHE) to construct and classically simulate the classical-quantum circuits that are needed for the quantum homomorphic encryption simulation.\\

As an example of application of the QHE scheme, we are interested on quantum walks. These algorithms were born from the quantization of classical random walks. They were first proposed in the discrete time version \cite{QRW}, and later using a continuous time \cite{Trees}. These walks have given rise to a wide variety of algorithms for problems such us triangle finding \cite{Triangles}, element distinctness \cite{ED} and quantum search \cite{QRW_Search}. Moreover, they can simulate a lot of physical systems \cite{Portugal}. An important quantum walk in discrete time is the one introduced by Szegedy \cite{Szegedy} as a generalization of the Grover algorithm \cite{Grover}. In contrast to other approaches, which are only useful for regular graphs, this quantum walk can quantize a general Markov chain. Thus, it can be used for any arbitrary weighted graph. Moreover, it has been shown to be useful for problems of optimization \cite{Lemieux,Qfold,QMS,GWQMA,qBIRD}, classification \cite{Paparo1,Paparo2,APR}, quantum search \cite{Portugal,Search_walk,Searchrank,S_queries} and machine learning \cite{Paparo3}. Although a previous study of quantum walks with QHE was done \cite{Rhode}, it was focused in the context of Boson sampling using a multi walker, and differs radically from the usual quantum walks studied in the literature \cite{Portugal}. Moreover, the QHE scheme applied was very limited, not being universal. As a remark, the graphs we have studied serve as toy models for testing Szegedy quantum walk simulation. Although these graphs are limited, they still have some applications in search algorithms \cite{Graph-phased,Portugal}. Our results are intended as a proof of concept for QHE simulation, and we plan to expand this work to include more complex graphs soon.\\

Recently, another important type of walk algorithm combining classical and quantum features, denoted as semiclassical walk, was developed \cite{Semiclassical}. They have been studied in the context of searching and ranking nodes in graphs, showing advantages with respect both classical and quantum algorithms \cite{Semiclassical,Randomized}. Furthermore, they have been applied to blockchain technologies \cite{Blockchain}. From a practical point of view, these walk algorithms are based on repeated measurements of the quantum walker's position at fixed time intervals. This fact could be crucial in algorithms that require knowing the position not only at the end of the quantum walk. After each measurement part of the system must be reset, so that the information of the qubits must be erased and the system must be reinitialized either with quantum or classically controlled gates able to read the previous measurement result. In order to perform a QHE scheme for semiclassical walks, the rules for the homomorphic evaluation of the reset and intermediate measurement operations must also been taken into account. We have also provided the rules for these operations in this work.\\

We hereby summarize briefly some of our main results:
\begin{enumerate}[i)]
    \item We reformulate Liang's schemes using classical-quantum circuits for the decryption procedure. The rules for the reset and intermediate measurement operations have also been included. This allows the QHE implementation of semiclassical walks.
    \item We examine the $T/T^\dagger$-gate complexity of the circuits for Szegedy quantum walk on different graphs: cyclic graphs, complete graphs and bipartite graphs. We show that all the circuits we have considered can be implemented homomorphically in an efficient manner. 
    \item  We simulate a homomorphic Szegedy quantum walk on the bipartite graph for six qubits using Qiskit. The correct results are always obtained after decryption, independently of the initial key used in encryption. To check that the results are indeed the correct ones, we have used the python library SQUWALS \cite{Squwals} to simulate the walk algorithm deterministically. After that we simulate a homomorphic semiclassical walk on a cycle graph that contains six qubits as well. Once more, SQUWALS was used to simulate the walk deterministically in order to check that the decrypted results of the QHE simulation are correct. The positive results obtained from both simulations show the correct functioning of the QHE scheme.
    \item We introduce the library CQC-QHE to construct and classically simulate the classical-quantum circuits required for quantum homomorphic encryption simulation scenarios.
\end{enumerate}

This paper is organized as follows. In Section \ref{QHE} Liang's schemes using classical-quantum circuits for the decryption procedure are reinterpreted. In Section \ref{Szegedy} we review the formulation of Szegedy quantum and semiclassical walk. In Section \ref{Circuits} the $T/T^{\dagger}$-gate complexity of different walks is analyzed to show the efficiency of the homomorphic implementation. In Section \ref{Simulation} the simulations performed in Qiskit are shown. Finally, we summarize and conclude in Section \ref{Conclusions}. In Appendix \ref{A:gates} the quantum gates used in the scheme are reviewed. In Appendix \ref{A:gate_teleportation}, gate teleportation is reviewed. In the supplementary material  the implementation of classical gates in Qiskit, the details of the simplified simulator and a proof of correctness for the update operator circuit of the bipartite graph are shown.
\newpage

\section{Quantum homomorphic encryption}\label{QHE}

In this section we describe a QHE scheme based on gate teleportation similar to Liang's scheme. However, we do it in a more practical manner. For a rigorous mathematical description we refer to the original work \cite{Liang}. Moreover, we will show how this scheme can be easily implemented in terms of classical-quantum circuits.\\

\subsection{Scheme description}

Let us consider a quantum algorithm as the process of applying a quantum circuit representing an unitary operator $U$ to an initial state $\left|\alpha\right>$, in order to obtain the quantum state $\left|\beta\right> = U\left|\alpha\right>$. Client is able to prepare $\left|\alpha\right>$, and Server is in charge of applying the quantum circuit of the algorithm. Client does not want Server to be able to obtain information about $\left|\alpha\right>$, so the quantum algorithm is performed using quantum homomorphic encryption. The QHE scheme used in this work consists on three main procedures, two of them performed by Client and one by Server. Moreover, the quantum circuit performing $U$ must be decomposed in Clifford+$T$ gates from the set $\mathcal{G} = \lbrace{X,Z,H,S,S^\dagger,CNOT,T,T^\dagger\rbrace}$. The mathematical expression of these gates is shown in Appendix \ref{A:gates}.\\

{\bfseries Step 1:} Client initializes a $n$-qubit system in the desired initial state $\left|\alpha\right>$ of the quantum algorithm. The qubits are denoted as $q_i$ for $i=1,..,n$. Before sending the qubits to Server, who is going to run the algorithm, Client must encrypt the system with quantum one-time pad (QOTP) encryption \cite{One_time_pad}. For doing so, Client generates randomly a secret key composed of two bit strings of length $n$. These two bit strings are stored in classical-bitstring variables. Let us denote $x$ and $z$ to these classical-bitstring variables, so that:
\begin{equation}
x = x_1x_2...x_n,
\end{equation}
\begin{equation}
z = z_1z_2...z_n.
\end{equation}
In order to encrypt the qubits, Client applies to each qubit $q_i$ Pauli gates $X$ and $Z$ depending on the classical bits $x_i$ and $z_i$, which at this moment contain the initial secret key. Thus, Client prepares the encrypted state
\begin{equation}
\left|\alpha^{enc}\right> = \left[\bigotimes_{i=1}^n X_i^{x_i}Z_i^{z_i}\right] \left|\alpha\right>,
\end{equation}
and sends the qubits to Server.\\

{\bfseries Step 2:} The quantum circuit that Server must run can be read sequentially gate by gate. For each gate $g$ in the original circuit, Server applies its corresponding homomorphic evaluation scheme. Whereas this evaluation consists trivially of applying gate $g$ for Clifford gates, it is quite more complicated for $T/T^\dagger$ gates, as we will show below. Each gate evaluation produces a change in the encrypting key of the system, so that the values stored in classical bits $x_i$ and $z_i$ will have to be updated accordingly. For each gate $g$ (CNOT) applied over qubit $q_i$ (qubits $q_i$ and $q_j$), there exists a key-updating function, denoted as $f_{g,i}$ ($f_{C,ij}$), that tells how the associated classical bits $x_i$ and $z_i$ must be updated. Thus, Server generates a sequence with the key updating functions accordingly to the sequence of gates in the quantum circuit. Finally, Server sends the qubits with the state $\left|\beta^{enc}\right>$ to Client, along with the sequence of key-updating functions.\\

{\bfseries Step 3:} For each function $f_{g,i}$ or $f_{C,ij}$ in the sequence of key-updating functions, Client performs key updating of classical bits:
\begin{equation}
x_i,z_i \leftarrow f_{g,i}(x_i,z_i),
\end{equation}
\begin{equation}
(x_i,z_i),(x_j,z_j) \leftarrow f_{C,ij}((x_i,z_i),(x_j,z_j)).
\end{equation}
At the end of the sequence, these classical bits contain the final encryption key. Thus, Client can obtain the decrypted final state applying QOTP decryption with these bits:
\begin{equation}
\left|\beta\right> = \left[\bigotimes_{i=1}^n X_i^{x_i}Z_i^{z_i}\right] \left|\beta^{enc}\right>.
\end{equation}

In principle, obtaining the final decrypted state $\left|\beta\right>$ is what Client needs if he/she wants to use it as the initial state for other algorithm, or in order to perform measurements in other basis different from the computational one. However, in most cases Client wants to measure the final state $\left|\beta\right>$ in the computational basis. In this case, Client can measure directly $\left|\beta^{enc}\right>$, so that the classical bitstring obtained as a result is encrypted with the key stored in classical bits $x_i$. Thus, Client can decrypt directly the classical result with this key using classical XOR operations instead of applying quantum gates.

\subsection{Evaluation schemes and key-updating functions for Clifford+$T$ gates}

So far we have only described our QHE scheme in an abstract sense. Now we show the explicit form of the evaluation schemes and key-updating functions for each quantum gate. In the following, we use characters $a$ and $b$ to denote constant values of bits, whereas we use $x$ and $z$ to denote classical-bit variables whose value can change.\\

Let $\left|\phi\right>$ be an arbitrary quantum state, as for example any intermediate state of the quantum algorithm. For single qubit gates, for the sake of simplicity let us suppose that they act on the first qubit of $\left|\phi\right>$, which is encrypted with $x_1=a$ and $z_1=b$ in the first qubit $q_1$, so that we have $X_1^{a}Z_1^{b}\left|\phi\right>$. We need to examine how the encrypting key changes after applying each of the quantum gates. Let us start with Clifford gates, and in particular the $X$ gate. We have, up to a global phase:
\begin{equation}
X_1 X_1^aZ_1^b\left|\phi\right> = X_1^aZ_1^b X_1 \left|\phi\right>,
\end{equation}
so after applying the $X$ gate over qubit $q_1$, the new state $X_1\left|\phi\right>$ is encrypted with $x_1=a$ and $z_1=b$. Thus, the classical bits have the same value before and after applying the quantum gate. If we applied a $Z$ gate instead, we would have:
\begin{equation}
Z_1 X_1^aZ_1^b\left|\phi\right> = X_1^aZ_1^b Z_1 \left|\phi\right>,
\end{equation}
so again the encrypting key is unchanged. Thus, the key-updating function for both gates is trivially the identity:
\begin{equation}
f_{X,i}(x_i,z_i) = f_{Z,i}(x_i,z_i) = (x_i,z_i).
\end{equation}
Now let us apply a Hadamard $H$ gate, so that
\begin{equation}
H_1 X_1^aZ_1^b\left|\phi\right> = X_1^bZ_1^a H_1 \left|\phi\right>.
\end{equation}
In this case the encrypting key changes, so that classical bits must be updated to $x_1 = b$ and $z_1 = a$. Thus, the key-updating function is
\begin{equation}
f_{H,i}(x_i,z_i) = (z_i,x_i).
\end{equation}
Finally, for $S$ and $S^\dagger$ gates we have
\begin{equation}
S_1 X_1^aZ_1^b\left|\phi\right> = X_1^aZ_1^{a \oplus b} S_1 \left|\phi\right>,
\end{equation}
\begin{equation}
S_1^\dagger X_1^aZ_1^b\left|\phi\right> = X_1^aZ_1^{a \oplus b} S_1^\dagger \left|\phi\right>,
\end{equation}
so in both cases the bits are updated to $x_1 = a$ and $z_1 = a \oplus b$. Then, both gates share the same key-updating function:
\begin{equation}
f_{S,i}(x_i,z_i) = f_{S^\dagger,i}(x_i,z_i) = (x_i,x_i \oplus z_i).
\end{equation}

For the $CNOT$ gate, let us suppose that the quantum state is encrypted with $x_1 = a_1$, $z_1 = b_1$, $x_2 = a_2$ and $z_2 = b_2$. We apply the $CNOT$ gate using qubit $q_1$ as control qubit and $q_2$ as target qubit. Then
\begin{equation}
CNOT_{12} X_1^{a_1}Z_1^{b_1}X_2^{a_2}Z_2^{b_2}\left|\phi\right> = X_1^{a_1}Z_1^{b_1 \oplus b_2}X_2^{a_1 \oplus a_2}Z_2^{b_2} CNOT_{12} \left|\phi\right>.
\end{equation}
In this case the classical bits must be updated to $x_1 = a_1$, $z_1 = b_1 \oplus b_2$, $x_2 = a_1 \oplus a_2$, $z_2 = b_2$, so that the key-updating function is given by
\begin{equation}
f_{C,ij}((x_i,z_i),(x_j,z_j)) = ((x_i,z_i \oplus z_j),(x_i \oplus x_j,z_j)).
\end{equation}

\begin{figure}[hbpt]
	\centering
	\includegraphics[scale=1]{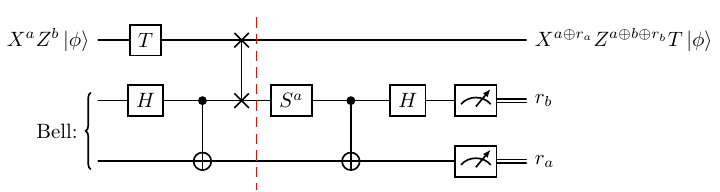}
	\caption{Homomorphic evaluation scheme for a $T$ gate using gate teleportation. Besides the $T$ gate, a swap gate, Clifford gates and measurements are performed. The part on the left of the red line is performed by Server, and the part on the right by Client once Server has finished running the quantum algorithm.}
	\label{F:T-evaluation}
\end{figure}

So far we have obtained the key-updating functions of Clifford gates. Since these gates are Clifford, when applied to any state encrypted with $X$ and $Z$ gates, they produce another state also encrypted with $X$ and $Z$ gates. Thus, the homomorphic evaluation scheme for these gates corresponds trivially on applying these gates. However, for non-Clifford gates this is not the case. Let us apply a $T$ gate over an encrypted state in the first qubit:
\begin{equation}
T_1 X_1^aZ_1^b\left|\phi\right> = (S_1^\dagger)^a X_1^aZ_1^{a \oplus b} T_1 \left|\phi\right>.
\end{equation}
The desired result $T\left|\phi\right>$ is also encrypted by $X$ and $Z$ gates, but there is an additional phase error, so that it is also encrypted by a phase gate $(S^\dagger)^a$. The bit value $a$ is given by the encrypting-key bit $x_1$ before the application of the $T$ gate. However, this depends in turn on the initial encrypting key, which Server does not know. Thus, Server is unable to correct this error by itself. Since this QHE scheme is non-interactive, Server cannot send the qubit to Client so that Client corrects the phase error until the end of the algorithm. For this reason, an evaluation scheme that makes use of gate teleportation was introduced \cite{Liang} (see appendix \ref{A:gate_teleportation}) in order to teleport a $S^a$ gate to the qubit. In Figure \ref{F:T-evaluation} we show the whole procedure. Server creates a Bell state in two ancilla qubits (denoted as Bell) using a Hadamard and a $CNOT$ gate. After that, Server applies a swap gate between the problem qubit and the first qubit of the Bell register. At this point, Server would send the two ancilla qubits to Client, and Client would perform a $S^a$-rotated Bell measurement in order to teleport the gate $S^a$ applied to the problem qubit. Thus, the qubit of Server ends up with the phase error corrected.\\

There are two differences with the actual gate teleportation protocol. In gate teleportation Server would have to apply $X^{r_{a}}$ and $Z^{r_{b}}$ in order to obtain the correct state after teleportation. However, in this case Server does not, and these two new bits, known only by Client, are considered part of the key updating. The second difference is that due to the principle of deferred measurements \cite{Nielsen}, Client can postpone these measurements until Server has finished running the quantum circuit. Thus, Server can wait until the end to send the ancilla qubits along with the main qubits of the system. Taking all this into account, the evaluation scheme for a $T$ gate corresponds to the part before the red line in Figure \ref{F:T-evaluation}, and the key-updating functions is given by:
\begin{equation}
f_{T,i;l}(x_i,z_i) = (x_i \oplus r_{a_l},x_i \oplus z_i \oplus r_{b_l}).
\end{equation}
Let $\nT$ be the number of $T/T^\dagger$ in the quantum circuit of the algorithm. For each of these gates, we need two different ancilla qubits, so we have in the end $\nT$ Bell registers. The new subscript $l = 1,...,\nT$ in the key-updating function makes reference to which Bell register Client must measure in order to obtain classical bits $r_{a_l}$ and $r_{b_l}$. Thus, when the function $f_{T,i;l}$ is read in the sequence of key-updating functions in Step 3, Client must also perform the quantum operations at the right of the red line in Figure \ref{F:T-evaluation} apart from the classical XOR operations. Thus, it can be thought as a kind of classical-quantum key-updating function.\\

Finally, for a $T^\dagger$ gate we have
\begin{equation}
T_1^\dagger X_1^aZ^b_1\left|\phi\right> = S_1^a X_1^aZ_1^{a \oplus b} T_1^\dagger \left|\phi\right>.
\end{equation}
If we use the same evaluation scheme as with the $T$ gate, in this case the teleported $S^a$ with the $S^a$ error produces $Z^a$, which cancels the $a$ superscript in the $Z$ gate, so the key updating function is simpler:
\begin{equation}
f_{T^\dagger,i;l}(x_i,z_i) = (x_i \oplus r_{a_l},z_i \oplus r_{b_l}).
\end{equation}

\subsection{Circuit representation}\label{QHE_circuits}

As we have seen above, the evaluation schemes that Server must apply for each quantum gate of the original circuit algorithm consist of applying quantum gates too, so the whole procedure of Step 2 can be represented by a quantum circuit. In the case of the decryption process of Client in Step 3 we also want a circuit representation. Each key-updating function of Clifford gates consists of applying XOR or swap operations, and this can be achieved with classical $CNOT$ and swap gates acting over the classical bits $x_i$ and $z_i$. Thus, we can represent the key-updating process with classical circuits. For $T/T^\dagger$ gates we can think of the key-updating function as a classical-quantum procedure, so we can represent it with a circuit that contains both classical and quantum gates. Thus, the whole procedure of Step 3 can be represented with a classical-quantum circuit. In Table \ref{tab:server-client} we summarize the building blocks for the circuits that Server and Client must run.

\begin{table}[htbp]
\caption{Classical-quantum circuits for the evaluation schemes performed by Server and the key-updating function performed by Client, associated to each Clifford+$T$ gate. Quantum bits are represented by single lines, whereas classical bits are represented by double lines. Note that for $T/T^\dagger$ gates, Client applies a $S$ gate controlled by classical bit $x$. Since this bit stores the value $a$, Client ends up applying $S^a$. For each quantum Bell register, denoted as $q$Bell, there is associated a classical register $c$Bell with two bits. The qubits of the registers $q$Bell start in state $\left|0\right>$. Notice that there is a fundamental difference in the evaluation of $T/T^\dagger$ gates compared to Clifford gates, since the former uses both qubits and classical bits whereas the latter only requires classical bits.}
\makebox[10pt][c]{
\begin{tabular}{ccc}
    \toprule
    Gate  & Server & Client \\
    \midrule
    $X$   &
    \begin{quantikz}
	\lstick[1]{$X^aZ^b\left|\phi\right>$} &\gate[1]{X}& \rstick[1]{ $X^aZ^bX\left|\phi\right>$}
    \end{quantikz}
    &
    \begin{quantikz}
   	\lstick[1]{$x: a$} \setwiretype{c} && \rstick[1]{$a$}\\
   	\lstick[1]{$z: b$} \setwiretype{c} && \rstick[1]{$b$}
    \end{quantikz}\\
    \midrule
    $Z$   &
    \begin{quantikz}
   	\lstick[1]{$X^aZ^b\left|\phi\right>$} &\gate[1]{Z}& \rstick[1]{ $X^aZ^bZ\left|\phi\right>$}
    \end{quantikz}
    &
    \begin{quantikz}
    	\lstick[1]{$x: a$} \setwiretype{c} && \rstick[1]{$a$}\\
    	\lstick[1]{$z: b$} \setwiretype{c} && \rstick[1]{$b$}
    \end{quantikz}\\
    \midrule
    $H$   &
    \begin{quantikz}
   	\lstick[1]{$X^aZ^b\left|\phi\right>$} &\gate[1]{H}& \rstick[1]{ $X^bZ^aH\left|\phi\right>$}
    \end{quantikz}
    &
    \begin{quantikz}
	\lstick[1]{$x: a$} \setwiretype{c} & \swap{1}& \rstick[1]{$b$}\\
    \lstick[1]{$z: b$} \setwiretype{c} &\targX{}& \rstick[1]{$a$}
    \end{quantikz}\\
    \midrule
    $S$   &
    \begin{quantikz}
   	\lstick[1]{$X^aZ^b\left|\phi\right>$} &\gate[1]{S}& \rstick[1]{ $X^aZ^{a \oplus b}S\left|\phi\right>$}
    \end{quantikz}
    &
    \begin{quantikz}
	\lstick[1]{$x: a$} \setwiretype{c} & \ctrl{1}& \rstick[1]{$a$}\\
    \lstick[1]{$z: b$} \setwiretype{c} &\targ{}& \rstick[1]{$a \oplus b$}
    \end{quantikz}\\
    \midrule
    $S^\dagger$   &
    \begin{quantikz}
   	\lstick[1]{$X^aZ^b\left|\phi\right>$} &\gate[1]{S^\dagger}& \rstick[1]{ $X^aZ^{a \oplus b}S\left|\phi\right>$}
    \end{quantikz}
    &
    \begin{quantikz}
	\lstick[1]{$x: a$} \setwiretype{c} & \ctrl{1}& \rstick[1]{$a$}\\
    \lstick[1]{$z: b$} \setwiretype{c} &\targ{}& \rstick[1]{$a \oplus b$}
    \end{quantikz}\\
    \midrule
    $CNOT$ & \ \ \ \
    \begin{quantikz}
	\lstick[1]{$X^{a_1}Z^{b_1}$} &  \setwiretype{n} \lstick[3]{} & & &\ctrl{2} \setwiretype{q}&& \setwiretype{n} X^{a_1}Z^{b_1 \oplus b_2}&\rstick[3]{ $CNOT\left|\phi\right>$}\\
    &\setwiretype{n}&& \lstick{$\left|\phi\right>$}&\\
    \lstick[1]{$X^{a_2}Z^{b_2}$} & \setwiretype{n} &&&\targ{} \setwiretype{q}&& \setwiretype{n} X^{a_1 \oplus a_2}Z^{b_2}&
    \end{quantikz}&
    \begin{quantikz}
   	\lstick[1]{$x_1: a_1$} \setwiretype{c} &\ctrl{2}&& \rstick[1]{$a_1$}\\
   	\lstick[1]{$z_1: b_1$} \setwiretype{c} &&\targ{}& \rstick[1]{$b_1 \oplus b_2$}\\
        \lstick[1]{$x_2: a_2$} \setwiretype{c} &\targ{}&& \rstick[1]{$a_1 \oplus a_2$}\\
   	\lstick[1]{$z_2: b_2$} \setwiretype{c} &&\ctrl{-2}& \rstick[1]{$b_2$}
    \end{quantikz}\\
    \midrule
    $T$   &
    \begin{quantikz}
	\lstick[1]{$X^aZ^b\left|\phi\right>$} &\gate[1]{T}&&\swap[]{1} \slice{}& \rstick[1]{ $X^{a\oplus r_a}Z^{a\oplus b \oplus r_b}T\left|\phi\right>$}\\
    \lstick[2]{$q$Bell:}& \gate{H} & \ctrl{1} & \targX{}&\\
    &&\targ{}&&
    \end{quantikz}
    &
    \begin{quantikz}[row sep=6,column sep=6]
	\lstick[2]{$q$Bell:}& \gate{S} & \ctrl{1} & \gate{H}& \meter{} \wire[d][2]{c}\\
    &&\targ{}&\meter{} \wire[d][2]{c}\\
    \lstick[2]{$c$Bell:} \setwiretype{c} &&&&&&&\ctrl{3}&\rstick[1]{$r_b$}\\
    \setwiretype{c} &&&&&&\ctrl{1}&&\rstick[1]{$r_a$}\\
    \lstick[1]{$x: a$} \setwiretype{c} &\ctrl{-4}&&&&\ctrl{1}&\targ{}&& \rstick[1]{$a\oplus r_a$}\\
    \lstick[1]{$z: b$} \setwiretype{c} &&&&&\targ{}&&\targ{}& \rstick[1]{$a\oplus b \oplus r_b$}
    \end{quantikz}\\
    \midrule
    $T^\dagger$ &
    \begin{quantikz}
   	\lstick[1]{$X^aZ^b\left|\phi\right>$} &\gate[1]{T^\dagger}&&\swap[]{1} \slice{}& \rstick[1]{ $X^{a\oplus r_a}Z^{b\oplus r_b}T^\dagger\left|\phi\right>$}\\
   	\lstick[2]{$q$Bell:}& \gate{H} & \ctrl{1} & \targX{}&\\
   	&&\targ{}&&
    \end{quantikz}
    &
    \begin{quantikz}[row sep=6,column sep=6]
    	\lstick[2]{$q$Bell}& \gate{S} & \ctrl{1} & \gate{H}& \meter{} \wire[d][2]{c}\\
    	&&\targ{}&\meter{} \wire[d][2]{c}\\
    	\lstick[2]{$c$Bell:} \setwiretype{c} &&&&&&\ctrl{3}&\rstick[1]{$r_b$}\\
    	\setwiretype{c} &&&&&\ctrl{1}&&\rstick[1]{$r_a$}\\
    	\lstick[1]{$x: a$} \setwiretype{c} &\ctrl{-4}&&&&\targ{}&& \rstick[1]{$a\oplus r_a$}\\
    	\lstick[1]{$z: b$} \setwiretype{c} &&&&&&\targ{}& \rstick[1]{$b \oplus r_b$}
    \end{quantikz}\\
    \bottomrule
    \end{tabular}
}
\label{tab:server-client}
\end{table}

\begin{figure}[htbp]
	\centering
	\includegraphics[scale=1]{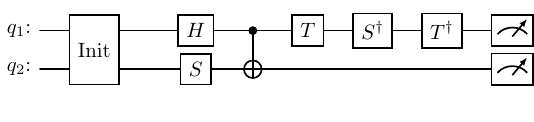}
	\caption{Example of a quantum algorithm with $2$ $T/T^\dagger$ gates and Clifford gates applied to two qubits. The block denoted by ``Init'' is an algorithm used by Client to create the initial state sent to Server. Its homomorphic implementation is not relevant since it is not performed by Server. The remaining Clifford+$T$ gates constitute Server's algorithm. The last step is measuring each qubit.}
	\label{F:naive_circuit}
\end{figure}

In Figure \ref{F:naive_circuit} we show an example of a quantum algorithm that we want to run with a QHE scheme. Client would create the initial state. Server would run the evaluation schemes associated to each gate, and generate the sequence of key-updating functions [$f_{H,1},f_{S,2},f_{C,12},f_{T,1;1},f_{S^\dagger,1},f_{T^\dagger,1;2}$], which is used by Client to update the encrypting key.

\begin{figure}[H]
\centering
\subfigure[]{\includegraphics[scale=1]{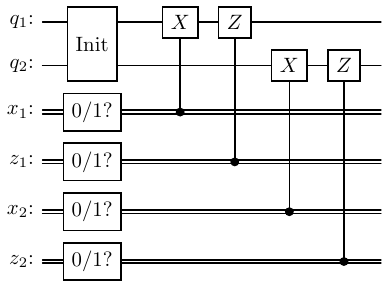}\label{F:client_circuit_1}}
\subfigure[]{\includegraphics[scale=1]{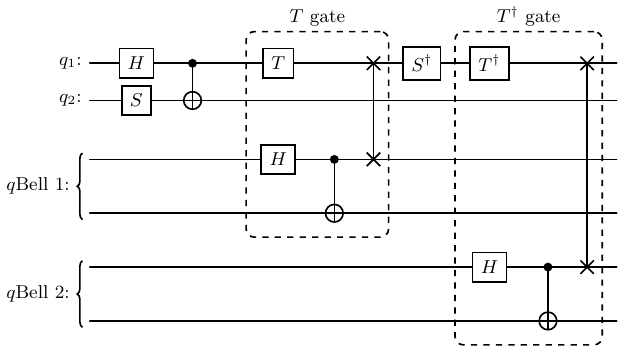}\label{F:server_circuit}}
\subfigure[]{\includegraphics[scale=1]{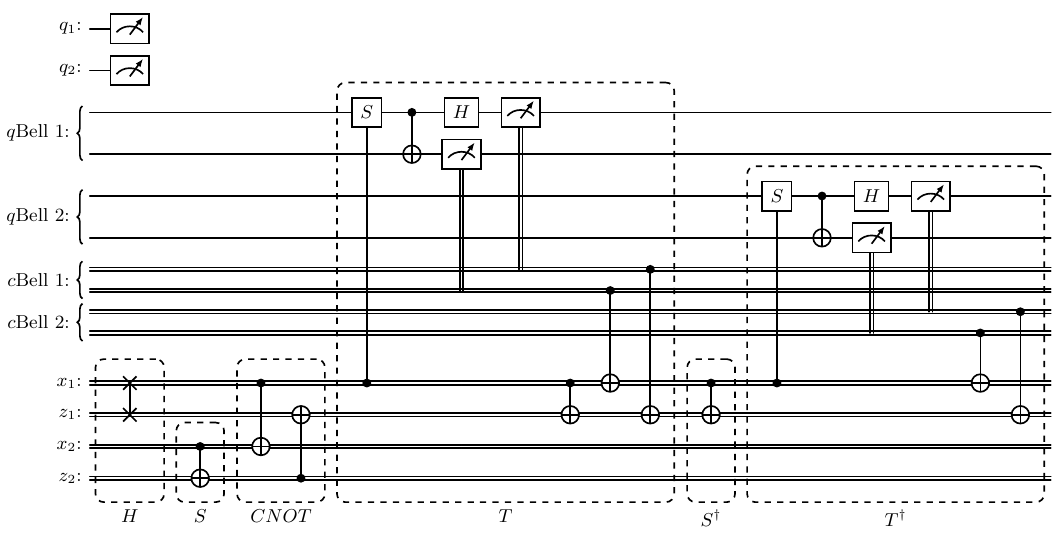}\label{F:client_circuit_2}}
\caption{Circuits for the three steps of the quantum homomorphic encryption scheme applied to the example in Figure \ref{F:naive_circuit}. a) Classical-quantum circuit for the initialization performed by Client in Step 1. Qubits $q_1$ and $q_2$ start in state $\left|0\right>$. b) Quantum circuit for the quantum algorithm performed by Server in Step 2. The qubits of the registers $q$Bell start in state $\left|0\right>$. c) Classical-quantum circuit for key-updating and measurement performed by Client in Step 3. Quantum bits are represented by single lines, whereas classical bits are represented by double lines.}
\label{F:classical-quantum-circuits}
\end{figure}

From this elementary but representative circuit, we can create the circuits that must be run in the three steps of the QHE scheme. In Figure \ref{F:client_circuit_1} we show the circuit that Client must run in order to initialize the system in Step 1. Note that before the encryption Client does not apply the evaluation scheme of the gates in order to create the initial state, so that the Init gate does not need to be compiled in the Clifford+$T$ gates of the universal set of gates $\mathcal{G}$. After creating the initial state, Client initializes the classical bits that store the encrypting key at random, and use classically-controlled quantum gates to encrypt the system. Since there are $n=2$ qubits, we have $2n=4$ classical bits. In Figure \ref{F:server_circuit} we show the homomorphic evaluation of the quantum gates of the algorithm. Since there are two $T/T^\dagger$ gates, we need two quantum Bell registers, thereby four ancilla qubits. Finally, in Figure \ref{F:client_circuit_2} we show the classical-quantum circuit that Client must run in order to obtain the final encrypting key. Note that in this example we measure the qubits of the main system, $q_1$ and $q_2$, before decryption, so that the classical results must be decrypted with classical bits $x_1$ and $x_2$. We will use this procedure in our simulations so that we can see the results before and after decrypting. However, we could instead apply $X$ and $Z$ gates controlled by their respective classical bits in order to obtain the final decrypted quantum state.

\subsection{Security and resources analysis of the QHE scheme}

The QHE scheme that we have introduced in this section is non-interactive and $\mathcal{F}$-homomorphic since it can be applied to any quantum circuit. QOTP is a protocol with perfect security, because if we select the encryption bitstrings at random and use them only once, applying this protocol to any arbitrary quantum state $\rho$ produces the totally mixed state \cite{One_time_pad}:
\begin{equation}\label{qotp_eq}
\frac{1}{2^{2n}} \sum_{a,b \in \lbrace{0,1\rbrace}^{n}} \left[\bigotimes_{i=1}^n X_i^{a_i}Z_i^{b_i}\right] \rho \left[\bigotimes_{i=1}^n X_i^{a_i}Z_i^{b_i}\right]^\dagger = \frac{\mathbbm{1}_{2^n}}{2^n}.
\end{equation}
Combining this with the fact that there is no interaction in the evaluation and decryption processes, then it turns out that this QHE has also perfect security.\\

Actually, our QHE scheme only differs from Liang's scheme \cite{Liang} in the form the key-updating functions are created. As it has been presented we do not have \nT-quasi-compactness since the number of key-updating functions scales linearly with the number of gates in the circuit, independently of being Clifford or $T/T^\dagger$ gates. This is so because we have an independent key-updating function per quantum gate. However, we can slightly modify the scheme so that we achieve $\nT$-quasi-compactness and it resembles more Liang's scheme.\\

Since all key-updating functions related to Clifford gates are deterministic functions not requiring quantum measurements, we can compose a sequence of Clifford key-updating functions into a single one denoted as $\widetilde{f}_l$. Thus, before the first $T/T^\dagger$ gate, between each of them and after the last one we have a single key-updating function $\widetilde{f}_l$, where $l = 1,...,\nT+1$. This function can be performed only by means of XOR operations. For each new bit value, this is obtained as the XOR sum of the $2n$ classical bits. In this sum each bit appears at most once, since the XOR is self-inverse. Thus, the number of classical operations that are needed for each of the $2n$ new bit values is at most $2n-1$, not scaling with the number of quantum gates that determine the composed function $\widetilde{f}_l$. For the key-updating functions associated with $T/T^\dagger$ gates we require at most $3$ XOR operations. Since there are $\nT$ functions of this class, the maximum number of classical operations for decrypting, denoted as $|O_{cl}|$, is
\begin{equation}
|O_{cl}|=(\nT+1)|\widetilde{f}_l| + 3\nT=(\nT+1)2n(2n-1) + 3\nT,
\end{equation}
where $|\widetilde{f}_l|$ denotes the maximum number of classical operations of $\widetilde{f}_l$.
Thus, the complexity scales as $\mathcal{O}(\nT)$ and the QHE scheme has $\nT$-quasi-compactness.\\

In the circuit example of Figure \ref{F:naive_circuit} the sequence of key-updating functions would be [$\widetilde{f}_1,f_{T,1;1},\widetilde{f}_2,f_{T^\dagger,1;2},\widetilde{f}_3$], where the composed functions are:
\begin{equation}
\widetilde{f_1}((x_1,z_1),(x_2,z_2)) = ((z_1,x_1 \oplus x_2 \oplus z_2),(x_2 \oplus z_1,x_2 \oplus z_2)),
\end{equation}
\begin{equation}
\widetilde{f_2}((x_1,z_1),(x_2,z_2)) = ((x_1,x_1 \oplus z_1),(x_2,z_2)),
\end{equation}
\begin{equation}
\widetilde{f_3}((x_1,z_1),(x_2,z_2)) = ((x_1,z_1),(x_2,z_2)).
\end{equation}
Whereas the first function is an ensemble of the key-updating functions of the $H$, $S$ and $CNOT$ gates, the second one corresponds directly to the function of the $S^\dagger$ gate applied to the first qubit, and the third one is the identity since there are no gates after the last $T/T^\dagger$ gate.\\

Since for each $T/T^\dagger$ gate we need two ancilla qubits and Client has to perform $\mathcal{O}(\nT)$ classical-quantum decrypting operations, the efficiency of the scheme is directly related to the number of $T/T^\dagger$ gates. Thus, any quantum algorithm can be implemented efficiently provided that $\nT$ grows as a polynomial with the size of the problem \cite{Liang}.\\

Although in the end we have described a $\nT$-quasi-compact scheme using the composite Clifford key-updating functions, they cannot be implemented in Qiskit due to the current limitations. Therefore, for simulation purposes, in Section \ref{Simulation} we will use the unmodified scheme with a function per gate. Nevertheless, the composition of Clifford gates are theoretically feasible, and we aim to incorporate them in the near future.\\

Finally, regarding the privacy of Server's circuit, note that in the case of having a key-updating function per gate, although Server would not append any function for $X$ and $Z$ gates since they are trivial, Client could still obtain much information about the sequence of gates in the circuit. Even using the composite Clifford functions, each Clifford subcircuit in the entire circuit (up to Pauli operators on some qubits) still needs to be known to the Client, although not necessarily in the original form. Therefore, the security of Server's circuit is compromised, as in Liang's scheme. Nevertheless, it is usually Client who proposes the circuit to Server, as for example for the quantum walks we study in this work, so that it is not necessary for Server to keep the circuit secret.

\section{Szegedy quantum and semiclassical walks}\label{Szegedy}

\subsection{Quantum walk formulation}

A classical random walk on a weighted graph with $N$ nodes corresponds to a Markov chain, with associated $N \times N$ transition matrix $G$. The matrix elements $G_{ji}$ provide the probabilities of the walker jumping from node $i$ to node $j$. A general quantization of this process is obtained by Szegedy quantum walk \cite{Szegedy,Notes}. In this quantum walk the Hilbert space is the span of all the vectors representing the $N \times N$ directed edges of the graph, i.e.,
\begin{equation}
\mathcal{H} = \text{span}\lbrace\left|i\right>_1\left|j\right>_2,\ i,j = 0,1,...,N-1\rbrace = \mathbb{C}^N \otimes \mathbb{C}^N,
\end{equation}
where the states with indexes $1$ and $2$ refer to the nodes on two copies of the original graph. Thus, the states are defined over two quantum registers, which can be reinterpreted as position and coin registers \cite{Semiclassical,Graph-phased}. By convention we count the nodes of the network, and therefore the matrix indexes, from $0$ to $N-1$. We define the vectors
\begin{equation}\label{psi_i}
	\left|\psi_i\right> := \left|i\right>_1 \otimes \sum_{k=0}^{N-1} \sqrt{G_{ki}}\left|k\right>_2,
\end{equation}
which are a superposition of the vectors representing the edges outgoing
from the $i^{th}$ vertex, whose coefficients are given by the square root of the $i^{th}$ column of the matrix $G$. From these vectors we define a projector operator onto the subspace generated by them:
\begin{equation}\label{projector}
	\Pi := \sum_{i=0}^{N-1} \left|\psi_i\right>\left<\psi_i\right|.
\end{equation}
The quantum walk operator $U_w$ is defined as
\begin{equation}\label{U}
	U_w := S_wR,
\end{equation}
where $R$ is a reflection over the subspace generated by the $\left|\psi_i\right>$ states,
\begin{equation}\label{reflection}
R := 2\Pi - \mathbbm{1},
\end{equation}
and $S_w$ is the swap operator between the two quantum registers, i.e.,
\begin{equation}\label{swap}
	S_w := \sum_{i,j=0}^{N-1} \left|i,j\right>\left<j,i\right|.
\end{equation}
The initial state of the system $\left|\psi(0)\right>$ is usually a linear combination of the $\left|\psi_i\right>$ states representing the initial probability distribution of the walker. After applying the unitary evolution $U$ a number $t$ of time steps, the position of the walker is usually obtained by measuring the first register. Thus, the probability distribution $p_q(t)$ is provided by the projection onto the computational basis of the first register.
\begin{equation}
\left[p_q(t)\right]_i = \left|\left|\tensor[_1]{\left<i\right|U_w^{t}\left|\psi(0)\right>}{}\right|\right|^2.
\end{equation}
There are algorithms where the second register is measured instead \cite{Paparo1,Paparo2,APR,Randomized}. However, without loss of generality in this work we will only consider measurements in the first register.

\subsection{Semiclassical walk formulation}

Semiclassical walks are a new kind of algorithms that combine both classical and quantum features \cite{Semiclassical}, which have demonstrated superior performance compared to both classical and quantum algorithms in ranking and search problems \cite{Randomized}. From a functional point of view in a quantum computer, these walks consist of repeated measurements of the walker position at regular intervals of time. There are two parameters to describe a semiclassical walk:
\begin{itemize}
\item Quantum time \cite{QT} $t_q$, which is the number of times we apply the walk unitary evolution operator $U_w$ between measurements.
\item Classical time $t_c$, which is the number of times that the scheme of quantum evolution and measurement is repeated.
\end{itemize}
In the case of the semiclassical Szegedy walk, there are two classes of algorithms, class I and class II, depending on whether register 1 or 2 is being measured respectively to obtain the information about the position of the walker. In this case we will only deal with the semiclassical walks of class I.\\

\begin{figure}[hbpt]
	\centering
	\includegraphics[scale=1]{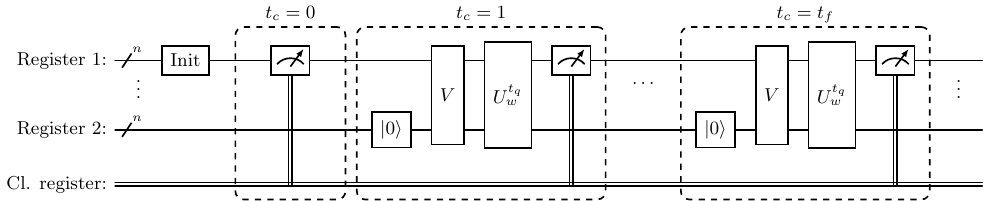}
	\caption{Quantum circuit for the semiclassical Szegedy walk. The walk is performed $t_f$ classical time steps, and at each classical step the quantum unitary evolution $U_w$ is applied $t_q$ times. After each classical step measurement in the first register and before the next quantum walk evolution, the second register is reset to $\left|0\right>_2$ and the update operator $\update$ is applied to create the new initial state $\left|\psi_{x}\right>$ depending on the result $x$ of the measurement.}
	\label{F:semiclassical_I}
\end{figure}

At each classical step $t_c$, the classical position of the walker $x_{t_c}$ is represented in the quantum system by its corresponding state $\left|\psi_{x_{t_c}}\right>$ in \eqref{psi_i}. In order to prepare this state, we need to introduce the update operator $\update$, which prepares the state $\left|\psi_i\right>$ from the computational basis of the first register, provided the second one is in state $\left|0\right>_2$ \cite{Q_circuits_2}. Thus:
\begin{equation}\label{update}
\update\left|i\right>_1\left|0\right>_2 = \left|\psi_i\right>.
\end{equation}
The implementation of a semiclassical walk in terms of quantum circuits is shown in Figure \ref{F:semiclassical_I}. First, an initial state is generated in the first register. This state represents the initial classical distribution of the walker at time step $t_c=0$. It can be constructed as an actual quantum superposition of the computational basis where all coefficients are positive or as a mixed state. After the measurement at $t_c=0$ the register results in the state $\left|x_0\right>_1$. From this point, each classical step $t_c > 0$ consists of four operations. The second register is reset to $\left|0\right>_2$, so that its previous information is erased and the system goes to state $\left|x_{t_c-1}\right>_1\left|0\right>_2$. The update operator creates the initial state of the step $\left|\psi_{x_{t_c-1}}\right>$. The quantum unitary evolution $U_w$ of Szegedy quantum walk is applied a number of times $t_q$, and finally the first register is measured to obtain the new classical position $x_{t_c}$. This process is repeated until the last classical step $t_f$.\\

From a formal point of view, semiclassical walks can be understood as classical walks where the transition matrices encode the quantum evolution. These are called semiclassical transition matrices, and are defined as:
\begin{equation}\label{G2}
	G^{(t_q)}_{ji} := \left|\left|\tensor[_1]{\left<j\right|U_w^{t_q}\left|\psi_i\right>}{}\right|\right|^2.
\end{equation}
Note that we actually have an entire family of semiclassical walks, parameterized by the quantum time $t_q$, and the classical time is the actual parameter that determines the time steps of the walker. The semiclassical matrix could be used to simulate the semiclassical walk deterministically in a classical computer, provided we have an efficient classical simulator of the quantum walk to obtain all the matrix elements \cite{Squwals}.

\subsection{QHE scheme with semiclassical framework}

For a homomorphic implementation of a semiclassical walk, Client would create the initial state representing the classical distribution, and Server would perform the rest of the circuit in Figure \ref{F:semiclassical_I}. In this case we notice that apart from usual quantum operators, which are expressible in Clifford+$T$ gates, we also have intermediate measurements and qubit reset operations. Thus, we need key-updating functions for this new operations. Although in the literature there are rules for final measurements and new qubits initialization \cite{Broadbent}, they are not considered as intermediate operations in a quantum circuit and do not consider the measured state of the system. Thus, we analyze deeper this case to check that the rules hold.\\

For the measurement operation, suppose we have a qubit that after being measured results in the state $\left|0\right>$ with probability $p_0$ and in the state $\left|1\right>$ with probability $p_1$. If we apply an $X$ gate these probabilities are exchanged. Then, if after measuring a qubit the resulting state is $\left|m\right>$, the encrypted state with $X^a$ would result in the state $X^a\left|m\right>$, where $m \in \lbrace{0,1\rbrace}$. Applying a $Z$ gate does not affect to the measurement probabilities, so it plays no role. Moreover, the resulting state is always an eigenstate of the $Z$ gate, so that without loss of generality we can consider that the resulting state is not encrypted by any $Z$ gate. Thus, the key-updating function associated to the measurement operator is
\begin{equation}
f_{M,i}(x_i,z_i) = (x_i,0).
\end{equation}
For the reset operation, regardless of the encrypting key of the qubit, the final state is always $\left|0\right>$, so it is not encrypted by any gate. Thus, the associated key-updating function is
\begin{equation}
f_{R,i}(x_i,z_i) = (0,0).
\end{equation}
In Table \ref{tab:server-client_sc} we show the circuits that Server and Client must apply for these two operations. Note that these operations behave as Clifford gates, so that their key-updating functions are deterministic and can be composed with Clifford key-updating functions in order to achieve the $\nT$-quasi-compactness.\\

\begin{table}[htbp]
\caption{Classical-quantum circuits for the evaluation schemes performed by Server and the key-updating function performed by Client, associated to measurement and reset operations. Quantum bits are represented by single lines, whereas classical bits are represented by double lines.}
\makebox[10pt][c]{
\begin{tabular}{ccc}
    \toprule
    Gate  & Server & Client \\
    \midrule
    \begin{quantikz}
	\meter{}
    \end{quantikz}   &
    \begin{quantikz}
	\lstick[1]{$X^aZ^b\left|\phi\right>$} &\meter{}& \rstick[1]{ $X^aZ^0\left|m\right>$}
    \end{quantikz}
    &
    \begin{quantikz}
   	\lstick[1]{$x: a$} \setwiretype{c} && \rstick[1]{$a$}\\
   	\lstick[1]{$z: b$} \setwiretype{c} &\gate{0}& \rstick[1]{$0$}
    \end{quantikz}\\
    \midrule
    $\left|0\right>$   &
    \begin{quantikz}
   	\lstick[1]{$X^aZ^b\left|\phi\right>$} &\gate[1]{\left|0\right>}& \rstick[1]{ $X^0Z^0\left|0\right>$}
    \end{quantikz}
    &
    \begin{quantikz}
    	\lstick[1]{$x: a$} \setwiretype{c} &\gate{0}& \rstick[1]{$0$}\\
    	\lstick[1]{$z: b$} \setwiretype{c} &\gate{0}& \rstick[1]{$0$}
    \end{quantikz}\\
    \bottomrule
    \end{tabular}
}
\label{tab:server-client_sc}
\end{table}

Finally, let us analyze the security of the algorithm. In this case, since the initial state represents a classical distribution, it is a linear combination of the computational basis states without relative phases or a mixed state. Then, all information can be obtained measuring only in the computational basis, whose results are invariant under the application of $Z$ operators. Thus, this state can only be encrypted using $X$ gates. Although the encrypting key is half of the usual one because it has only a single bitstring instead of two, the initial state also carries less information than an usual quantum state. Thus, we can prove that for states representing classical distributions, QOTP encryption with $X$ gates also provides perfect security. For a classical distribution state, it does not matter if it is a quantum superposition or a mixed state, so let us consider the mixed state
\begin{equation}
\rho_c = \sum_{i=0}^{N-1} p_i \left|i\right>\left<i\right|,
\end{equation}
where $p_i$ are the probabilities of the classical distribution that this state represents. It is trivial that $\rho_c = Z_i\rho_c Z_i^\dagger$ for any $Z$ gate acting over any qubit. Thus, if we apply all random keys with $X$ gates we obtain the totally mixed state again:
\begin{equation}\label{qotp_eq_sc}
\frac{1}{2^{n}} \sum_{a \in \lbrace{0,1\rbrace}^{n}} \left[\bigotimes_{i=1}^n X_i^{a_i}\right] \rho_c \left[\bigotimes_{i=1}^n X_i^{a_i}\right]^\dagger = \frac{1}{2^{2n}} \sum_{a,b \in \lbrace{0,1\rbrace}^{n}} \left[\bigotimes_{i=1}^n X_i^{a_i}Z_i^{b_i}\right] \rho_c \left[\bigotimes_{i=1}^n X_i^{a_i}Z_i^{b_i}\right]^\dagger = \frac{\mathbbm{1}_{2^n}}{2^n},
\end{equation}
where in the last step we have used equation \eqref{qotp_eq}.

\section{$T/T^\dagger$ count for Szegedy quantum circuits}\label{Circuits}

As we previously mentioned, a quantum algorithm can be efficiently implemented in a QHE scheme if the number of $T/T^\dagger$ gates scales as a polynomial with the size of the problem. For Szegedy quantum walk, the size of the problem is expressed in terms of the number of nodes of the graph $N$. The number of times steps of the quantum walk depends on the particular algorithm. However, as long as the algorithm is efficient, then the number of time steps will scale polynomially with $N$. For example for search problems in graphs this scales with the square root of $N$ \cite{Portugal,S_queries}. Thus, the efficiency of the implementation ultimately depends on the quantum circuit compilation of the unitary evolution operator $U_w$ in \eqref{U}.\\

\begin{figure}[hbpt]
	\centering
	\includegraphics[scale=1]{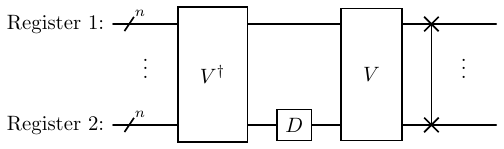}
	\caption{General quantum circuit for the unitary evolution operator $U_w$ of Szegedy quantum walk in terms of the update operator $\update$, a diagonal operator $D$ and the swap operator $S$.}
	\label{F:szegedy_circuit_general}
\end{figure}

To obtain a quantum circuit for Szegedy quantum unitary $U_w$, we need two registers of qubits. Each register needs to represent $N$ possible states for a graph with $N$ nodes. From now on we consider that each register has $n$ qubits, being $N = 2^n$, so that the total number of qubits for the circuit is $2n$. Szegedy quantum walk unitary evolution $U_w$ \eqref{U} consists of two operators: the swap $S_w$ and the reflection $R$. For a general circuit, we need to diagonalize the reflection operator \eqref{reflection} using the update operator $\update$ \cite{Q_circuits,Q_circuits_2} whose action was defined in \eqref{update}, such that:

\begin{equation}
\update^\dagger R \update = 2\sum_{i=0}^{N-1} \update^\dagger\left|\psi_i\right>\left<\psi_i\right|\update - \mathbbm{1} = 2\sum_{i=0}^{N-1} |i \rangle \raisebox{-0.5ex}{\small$_1$} \langle i| \otimes |0 \rangle \raisebox{-0.5ex}{\small$_2$} \langle 0| - \mathbbm{1} = \mathbbm{1}\otimes D,
\end{equation}
where
\begin{equation}
D = 2|0 \rangle \raisebox{-0.5ex}{\small$_2$} \langle 0| - \mathbbm{1}
\end{equation}
is a diagonal operator acting in the second register. Thus, the reflection $R$ can be decomposed as $R = \update(\mathbbm{1}\otimes D)\update^\dagger$. From this, we can construct a general circuit for the unitary operator $U_w$ of Szegedy quantum walk as shown in Figure \ref{F:szegedy_circuit_general}, where we have four operators. The swap operator $S_w$ is constructed by swap gates between the qubits of the first and second registers. Since the swap gates can be decomposed as three $CNOT$ gates \cite{Nielsen}, they do not contribute to the count of $T/T^\dagger$ gates.\\

\begin{figure}[hbpt]
	\centering
	\includegraphics[scale=1]{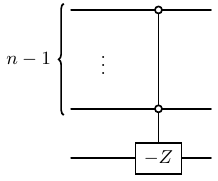}
	\raisebox{2cm}{=}
	\includegraphics[scale=1]{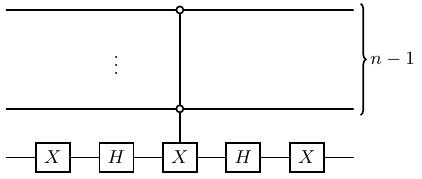}
	\caption{Quantum circuit for the diagonal operator $D$ in Figure \ref{F:szegedy_circuit_general}. It corresponds to a multi-controlled-$(-Z)$ gate. Since $-Z = XZX$ and $Z = HXH$, this gate can be implemented with a multi-controlled-$X$ gate.}
	\label{F:reflection}
\end{figure}

\begin{figure}[hbpt]
	\centering
	\subfigure[]{\includegraphics[scale=1]{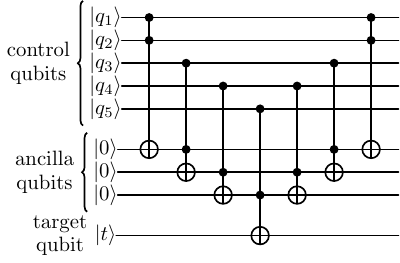}\label{F:multi-X}}
	\subfigure[]{\includegraphics[scale=1]{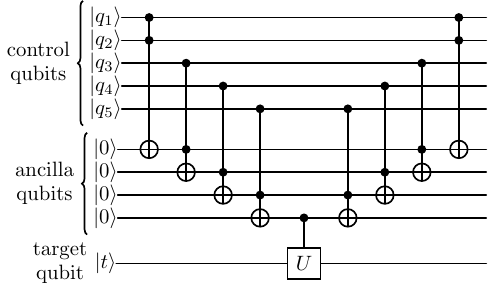}\label{F:multi-U}}
	\caption{a) Decomposition of a multi-controlled-$X$ operation that contains $n_c=5$ control qubits into $2n_c-3=7$ Toffoli gates using $n_c-2=3$ ancilla qubits. b) Decomposition of a multi-controlled-$U$ operation that contains $n_c=5$ control qubits into $2n_c-2=8$ Toffoli gates and a single-controlled-$U$ using $n_c-1=4$ ancilla qubits. These ancilla qubits must start in state $\left|0\right>$, and they end up also in state $\left|0\right>$. Thus, they can be reused for all multi-controlled gates in the circuit, and the total number of ancilla qubits only depends on the bigger multi-controlled gate, providing at most $n-2$ ancilla qubits for a circuit with $n$ qubits. Moreover, if they are provided by Client, they can be encrypted, as long as the initial state before encryption has these qubits in $\left|0\right>$.}
	\label{multicnot decomposition}
\end{figure}

\begin{figure}[hbpt]
	\centering
	\includegraphics[scale=1]{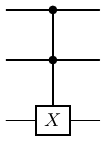}
	\raisebox{1cm}{=}
	\includegraphics[scale=1]{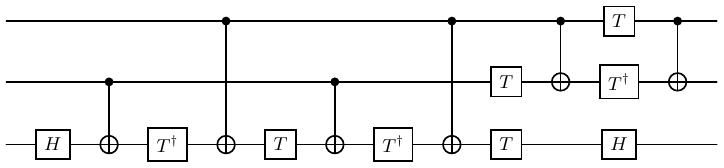}
	\caption{Decomposition of a Toffoli gate into Clifford+$T$ gates at a cost of $7$ $T/T^\dagger$ gates.}
	\label{F:Toffoli}
\end{figure}

The diagonal operator $D$ is a reflection around the state $\left|0\right>_2$ in the second register. This operator does not depend on the type of graph, but only on the number $n$ of qubits in each register. The circuit for this reflection is shown in Figure \ref{F:reflection}. Before going on, let us fix some notation about controlled gates. We denote as single-controlled-$U$ gate to a general $U$ gate being controlled by $n_c = 1$ control qubits, whereas we refer to multi-controlled-$U$ gate as a $U$ gate being controlled by $n_c>1$ control qubits. The $CNOT$ gate corresponds to a single-controlled-$X$ gate, whereas a Toffoli gate is a multi-controlled-$X$ gate with $n_c=2$ control qubits. Since the usual controlled gates that activate when the control qubits are in the state $\left|1\right>$ can be converted into controlled gates that activate when the control qubits are in the state $\left|0\right>$, surrounding them by Clifford $X$ gates, we do not worry about the particular controlling state for counting $T/T^\dagger$ gates. In the case of the operator $D$, we have a multi-controlled-$X$ gate with $n_c = n-1$ control qubits. This kind of gates can be decomposed into Toffoli gates as shown in Figure \ref{F:multi-X}, with the help of ancilla qubits. For $n_c$ control qubits, the decomposition yields $2n_c - 3$ Toffoli gates, so for $n_c = n-1$ we have $2n - 5$ Toffoli gates. Moreover, each Toffoli gate can be decomposed into Clifford+$T$ gates as shown in Figure \ref{F:Toffoli}, so that each of them provides $7$ $T/T^\dagger$ gates. The operator $D$ thus provides $14n - 35$ $T/T^\dagger$ gates. Since $n = \log_2(N)$ this operator can be efficiently implemented in a QHE scheme for any graph. Moreover, note that this formula is only valid for $n \geq 3$, since for $n \leq 2$ the number of $T/T^\dagger$ gates is $0$.\\

The update operator $\update^\dagger$ is the inverse of the operator $\update$. The circuit is constructed inverting the order of the gates and changing them by their inverse. For a circuit compiled in Clifford+$T$ gates, the inverse gates are also Clifford+$T$ gates, and the number of $T/T^\dagger$ gates is conserved. Thus, both operators contribute with the same number of $T/T^\dagger$ gates. Let $\nT_{\update}$ be the number of $T/T^\dagger$ gates provided by the operator $\update$, then the total number $L_U$ of $T/T^\dagger$ gates in the circuit for $U_w$ can be expressed as:

\begin{equation}\label{T_count_general}
    \nT_U = 14n-35 + 2\nT_{\update}.
\end{equation}
Thus, the efficiency of the implementation finally relies on the update operator $\update$. For a semiclassical walk this is also so, since the quantum circuit is composed of $\update$ and $U_w$ operators, and measurements and resets behave as Clifford gates.\\

\begin{figure}[hbpt]
	\centering
	\subfigure[]{\includegraphics[scale=0.375]{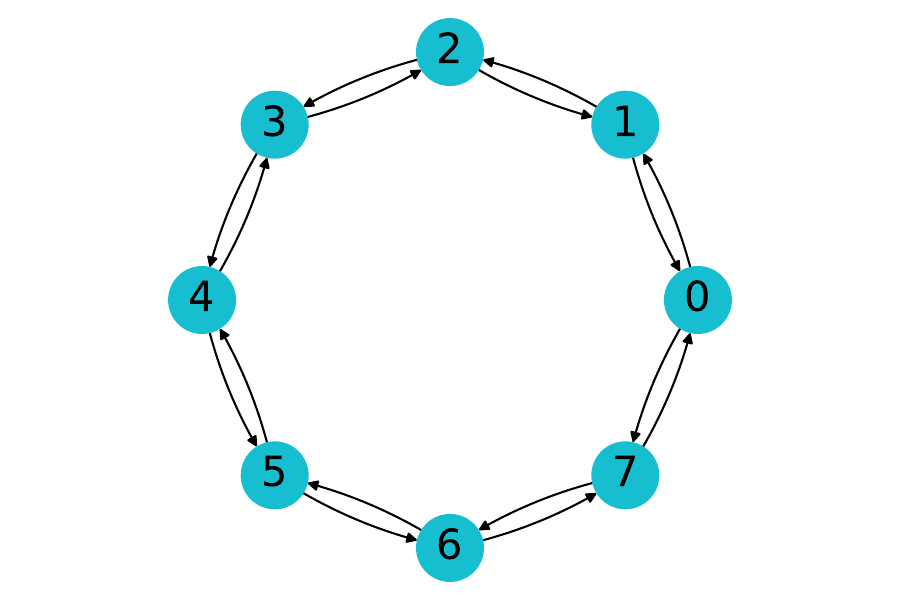}\label{F:cycle_graph}}
	\subfigure[]{\includegraphics[scale=0.375]{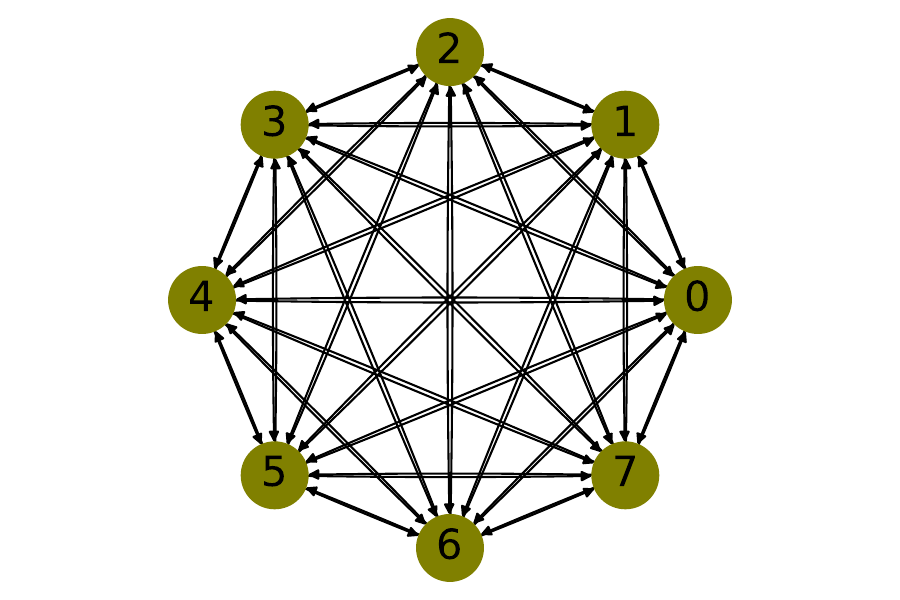}\label{F:complete_graph}}
	\subfigure[]{\includegraphics[scale=0.375]{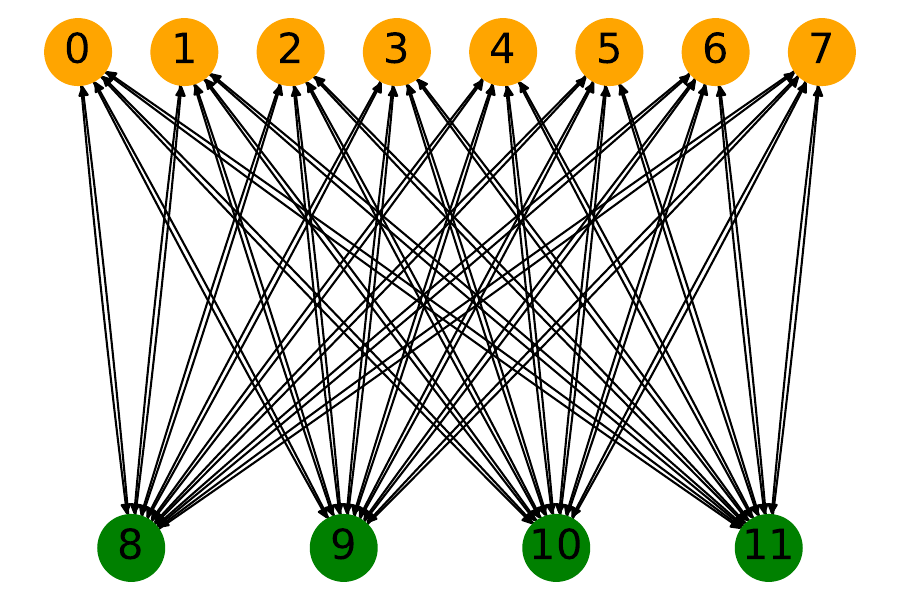}\label{F:bipartite_graph}}
	\caption{Representation of different types of graphs with symmetry. a) Cycle graph with $N=8$ nodes. b) Complete graph with $N=8$ nodes. c) Complete bipartite graph with a set of $N_1=8$ nodes shown in orange and another set of $N_2=4$ nodes shown in green. The graphs have been plotted using the python library NetworkX \cite{NetworkX}.}
	\label{F:graphs}
\end{figure}

The number of $T/T^\dagger$ gates of the update operator $\update$ will depend on the particular graph we are dealing with. In the following sections we show the decomposition for different types of graphs whose symmetry allows the construction of an efficient circuit. These graphs are shown in Figure \ref{F:graphs}. The main interest of these graphs is the quantum circuits they lead to, which are simple toy models for testing Szegedy quantum walk simulations, and also give insights for more complex graphs. Moreover, some of theme have applications in search algorithms \cite{Graph-phased}. Quantum circuits of these graphs scaling polylogarithmically in the number of gates have been proposed \cite{Q_circuits}. However, these circuits are provided in terms of multicontrolled gates and also general rotation gates, whose decomposition in Clifford+$T$ gates is not straightforward.\\

\subsection{Cyclic graph}

A cycle graph is shown in Figure \ref{F:cycle_graph}. It is formed by a closed line of $N$ nodes, where each node connects only to the front node and the back node. The transition matrix is
\begin{equation}
G_{ji} = \frac{1}{2}\delta_{i,j-1} + \frac{1}{2}\delta_{i,j+1}.
\end{equation}

\begin{figure}[hbpt]
	\centering
	\includegraphics[scale=1]{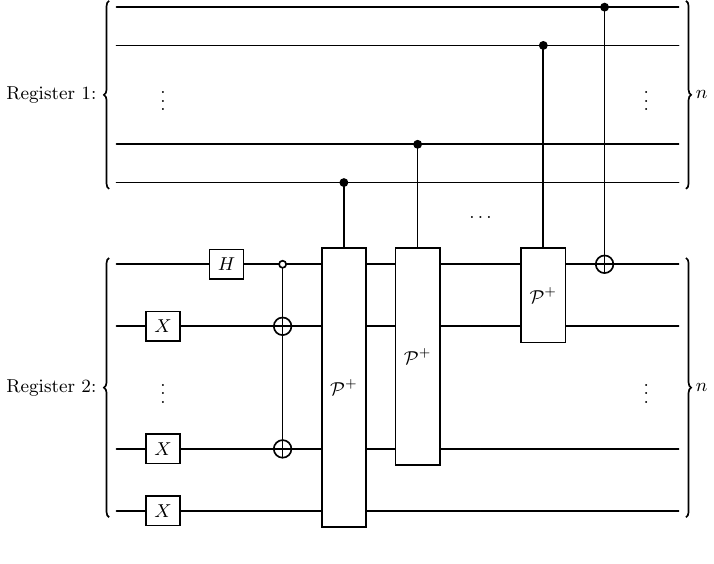}
	\caption{Quantum circuit for the update operator $\update$ for Szegedy quantum walk over a cycle graph.}
	\label{F:update_cycle}
\end{figure}

The quantum circuit for the update operator $\update$ \cite{Q_circuits} is shown in Figure \ref{F:update_cycle}. In this work we use big-endian ordering convention for qubits order, so that the left-most qubit in the tensor product of a multi-qubit state corresponds to the upper-most qubit of the circuit. The contribution of $T/T^\dagger$ gates comes from the permutation operators, denoted by $\adder$. These operators acting over $\adderqubits$ qubits add $1$ to each computational basis state:
\begin{equation}
\adder\left|x\right> = \left|x + 1 \ \text{mod} \ \adderqubits\right>.
\end{equation}

\begin{figure}[hbpt]
	\centering
	\includegraphics[scale=1]{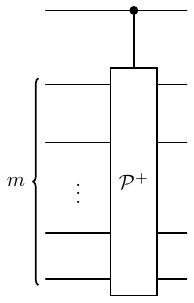}
	\raisebox{2.4cm}{=}
	\raisebox{0.04cm}{\includegraphics[scale=1]{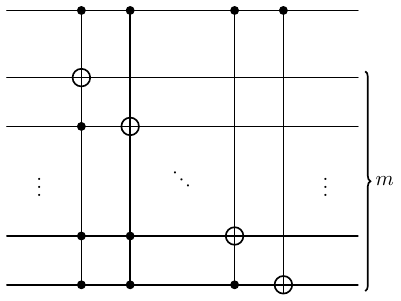}}
	\caption{Quantum circuit for a controlled-$\adder$ operator acting over $\adderqubits$ qubits.}
	\label{F:+1}
\end{figure}

The quantum circuit for a controlled-$\adder$ operator acting over $\adderqubits$ qubits is shown in Figure \ref{F:+1}. We have multi-controlled-$X$ gates from $n_c = 2$ control qubits to $n_c = \adderqubits$ control qubits. Again, we can decompose these gates into $2n_c-3$ Toffoli gates using the decomposition of Figure \ref{F:multi-X}. Each Toffoli can be decomposed in $7$ $T/T^\dagger$ gates, so that the number of $T/T^\dagger$ gates for a controlled-$\adder$ operator is:
\begin{equation}
L_{\adder}=7\sum_{n_c=2}^{\adderqubits} (2n_c-3) = 7(\adderqubits^2 - 2\adderqubits + 1).
\end{equation}
In the circuit for the update operator $\update$ we have controlled-$\adder$ operators from $\adderqubits=2$ to $\adderqubits=n$ target qubits. Thus, adding the number of $T/T^\dagger$ gates for each one we obtain the number of $T/T^\dagger$ gates for the update operator $\update$:
\begin{equation}
L_{\update}= 7\sum_{\adderqubits=2}^{n}(\adderqubits^2-2\adderqubits+1)=\frac{7}{6}(2n^3-3n^2+n).
\end{equation}

Substituting in equation \eqref{T_count_general} we obtain the number of $T/T^\dagger$ gates for the unitary evolution operator $U_w$ in the case of the cycle graph:
\begin{equation}\label{L_U_cycle}
L_U = \frac{14}{3}n^3 - 7n^2 + \frac{49}{3}n - 35.
\end{equation}
Finally, we use that $n = \log_2(N)$ to express $L_U$ in terms of the number of nodes:
\begin{equation}
L_U = \frac{14}{3}\log_2^3(N) - 7\log_2^2(N) + \frac{49}{3}\log_2(N) - 35.
\end{equation}
As we can observe, in the asymptotic limit the number of $T/T^\dagger$ gates scales as $\mathcal{O}(\log_2^3(N))$, so that it scales polylogarithmically with the size of the problem, and thus this walk can be implemented efficiently in a QHE scheme.

\subsection{Complete graph}

A complete graph is shown in Figure \ref{F:complete_graph}. It is formed by $N$ nodes connecting all the other nodes and without self-loops. The transition matrix is
\begin{equation}
G_{ji} = \frac{1}{N-1}(1 - \delta_{ij}).
\end{equation}

\begin{figure}[hbpt]
	\centering
	\includegraphics[scale=1]{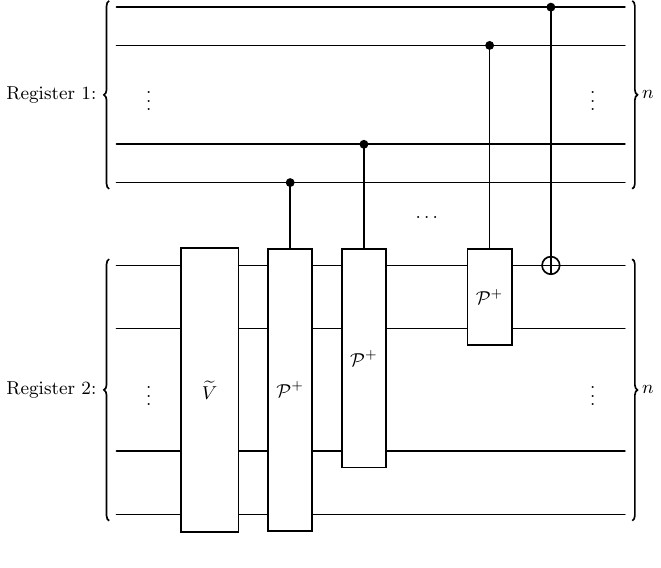}
	\caption{Quantum circuit for the update operator $\update$ for Szegedy quantum walk over a complete graph.}
	\label{F:update_complete}
\end{figure}

\begin{figure}[hbpt]
	\centering
	\includegraphics[scale=1]{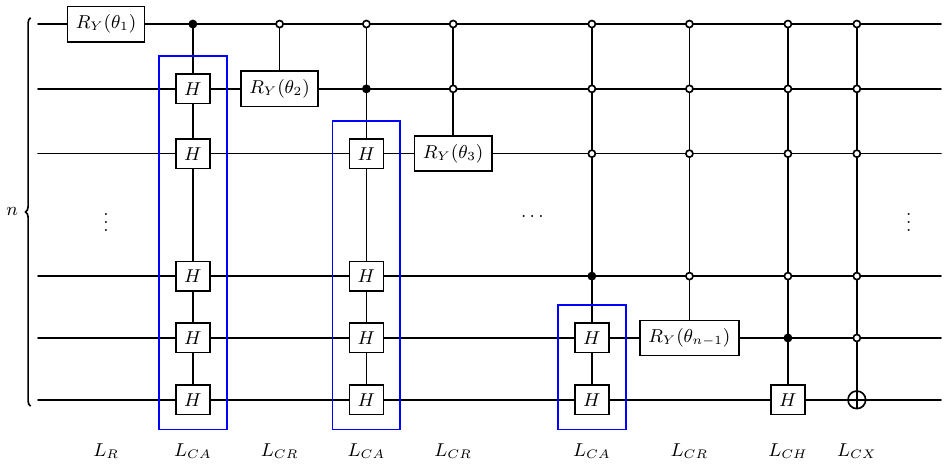}
	\caption{Quantum circuit for the operator $\widetilde{\update}$ in Figure \ref{F:update_complete}. The angles are given by $\theta_i = 2\arccos \left(\sqrt{(2^{n-i}-1)/(2^{n-i+1}-1)}\right)$. The formula of the $R_Y(\theta)$ gate is given in equation \eqref{ry}.}
	\label{F:update_complete_2}
\end{figure}

The quantum circuit for the update operator $\update$ \cite{Q_circuits} is shown in Figure \ref{F:update_complete}. The contribution to the count of $T/T^\dagger$ gates comes from the same $\adder$ operators as for the cycle graph, and a new contribution due to the operator $\widetilde{\update}$, whose quantum circuit is shown in Figure \ref{F:update_complete_2}. Thus, the number of $T/T^\dagger$ gates for the unitary evolution operator $U_w$ in the case of the complete graph is
\begin{equation}\label{L_complete}
L_U = \frac{14}{3}n^3 - 7n^2 + \frac{49}{3}n - 35 + 2L_{\widetilde{\update}}.
\end{equation}

In Figure \ref{F:update_complete_2} we have marked five different types of sources for the count of $T/T^\dagger$ gates, so that we can express this number for the $\widetilde{\update}$ operator as
\begin{equation}\label{L_K_2}
L_{\widetilde{\update}} = L_{CX} + L_{CH} + L_{CA} + L_{CR} + L_{R}.
\end{equation}
Since we are interested in the asymptotic scaling, we can assume that each register has $n \geq 3$ qubits, so that all these sources are present in the circuit.\\

The term $L_{CX}$ corresponds to the number of $T/T^\dagger$ gates due to the decomposition of the last multi-controlled-$X$ gate. It has $n_c=n-1$ control qubits. Using the decomposition in Toffoli gates of Figure \ref{F:multi-X}, and taking into account that each Toffoli provides $7$ $T/T^\dagger$ gates, we obtain that
\begin{equation}
L_{CX} = 7(2(n-1)-3) = 14n-35.
\end{equation}

The term $L_{CH}$ depends on the decomposition of the last multi-controlled-$H$ gate. Since the Hadamard gate can be expressed in the form $H = AXA^\dagger$, with $A = SHTHS^\dagger H$, the multi-controlled-$H$ gate can be transformed into a multi-controlled-$X$ gate surrounded by $A$ and $A^\dagger$ in the target qubits. Thus, this unitary transformation costs two $T/T^\dagger$ gates, which added to the contribution of the multi-controlled-$X$ gate results in:
\begin{equation}
L_{CH} = 7(2(n-1)-3) + 2 = 14n-33.
\end{equation}

The rest of controlled-$H$ gates come in the form of controlled arrays of $H$ gates, shown as blue boxes in Figure \ref{F:update_complete_2}. These controlled arrays contribute to the term $L_{CA}$. In this case we consider each array as a general unitary gate $U$, and apply a decomposition into $2n_c - 2$ Toffoli gates and an array of $H$ gates controlled by a single qubit, as shown in Figure \ref{F:multi-U}. On one hand, for the Toffoli gates we must take into account that there are different multi-controlled gates, from $n_c=1$ to $n_c=n-2$ control qubits, and that each Toffoli provides $7$ $T/T^\dagger$ gates. On the other hand, each single-controlled-array with $h$ Hadamard gates corresponds to $h$ single-controlled-$H$ gates, which again can be converted into Clifford $CNOT$ gates at a cost of two $T/T^\dagger$ gates. We need to consider the different arrays of $H$ gates, from $h=2$ to $h=n-1$ Hadamard gates. Thus, the total contribution of these gates can be expressed as
\begin{equation}
L_{CA} = 7\sum_{n_c=1}^{n-2} (2n_c-2) + 2\sum_{h=2}^{n-1}h = 8n^2 - 36n + 40.
\end{equation}

\begin{figure}[hbpt]
	\centering
	\includegraphics[scale=1]{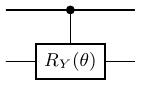}
	\raisebox{0.7cm}{=}
	\includegraphics[scale=1]{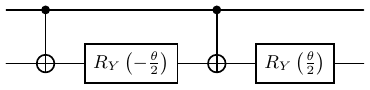}
	\caption{Decomposition of a single-controlled-$R_Y$ gate into two $CNOT$ gates and two $R_Y$ gates.}
	\label{F:cry}
\end{figure}

The term $L_{CR}$ corresponds to the number of $T/T^\dagger$ gates provided by the controlled-$R_Y$ gates. We can decompose these gates into $2n_c-2$ Toffoli gates and a single-controlled-$R_Y$ gate using again the decomposition of Figure \ref{F:multi-U}. The contribution due to the Toffoli gates is the same as for the term $L_{CA}$, since we have multi-controlled gates from $n_c=1$ to $n_c=n-2$ control qubits. Given the matrix form of a $R_Y$ gate as
\begin{equation}\label{ry}
R_Y(\theta) = 
\left(\begin{array}{cc}
	\cos\left(\frac{\theta}{2}\right) & -\sin\left(\frac{\theta}{2}\right) \\
	\\
	\sin\left(\frac{\theta}{2}\right) & \cos\left(\frac{\theta}{2}\right)
\end{array}\right),
\end{equation}
we have that $R_Y(\theta) = XR_Y(-\theta/2)XR_Y(\theta/2)$, so that a single-controlled-$R_Y$ gate can be decomposed into two $R_Y$ gates as shown in Figure \ref{F:cry}. Thus, the term $L_{CR}$ can be expressed as
\begin{equation}
L_{CR} = 7\sum_{n_c=1}^{n-2} (2n_c-2) + (n-2)2L_{R} = 7n^2 - 35n + 42 + (2n-4)L_{R},
\end{equation}
where $L_{R}$ is the number of $T/T^\dagger$ gates needed to compile a $R_Y$ gate, which so far we consider does not depend on the angle $\theta$.\\

Adding all the contributions in \eqref{L_K_2} we obtain the number of $T/T^\dagger$ gates for the operator $\widetilde{\update}$:
\begin{equation}
L_{\widetilde{\update}} = 15n^2 - 43n + 14 + (2n-3)L_{R}.
\end{equation}
Substituting in \eqref{L_complete} we obtain the number of $T/T^\dagger$ gates of the unitary evolution operator $U_w$:
\begin{equation}
L_U = \frac{14}{3}n^3 + 23n^2 - \frac{209}{3}n - 7 + (4n-6)L_{R},
\end{equation}
which in terms of the number of nodes of the graph ends up as
\begin{equation}\label{L_complete_2}
L_U = \frac{14}{3}\log_2^3(N) + 23\log_2^2(N) - \frac{209}{3}\log_2(N) - 7 + (4\log_2(N)-6)L_{R}.
\end{equation}
Assuming that a $R_Y$ can be implemented efficiently, with a scaling slower than the cube of the logarithm of the number of nodes, then the number of $T/T^\dagger$ gates scales as $\mathcal{O}(\log_2^3(N))$ and the unitary evolution $U_w$ can be implemented efficiently in a QHE scheme.\\

Solovay-Kitaev theorem states that any single qubit gate $U$ can be implemented in Clifford+$T$ gates with $\mathcal{O}(\log^c(1/\varepsilon))$ gates, where $c$ is a small constant approximately equal to $2$ and $\varepsilon$ is the error of the approximation using a finite number of gates \cite{Nielsen}:
\begin{equation}\label{Rz_error}
\varepsilon \geq \Vert \bar{U}-U \Vert,
\end{equation}
where $\bar{U}$ is the approximated gate. Thus, $R_Y$ gates can be implemented in principle efficiently for any angle $\theta$. Moreover, the $T/T^\dagger$ cost of implementing a $R_Y$ is the same as for a $R_Z$ gate since $R_Y(\theta) = SHR_Z(\theta)HS^\dagger$, and there are more efficient algorithms for decomposing a $R_Z$ in $T/T^\dagger$ gates. The algorithm \cite{Rz_decomposition} claims that any $R_Z$ gate needs in the worst case a number of $T/T^\dagger$ gates
\begin{equation}
L_{R} = 4\log_2(1/\varepsilon)+O(\log(\log(1/\varepsilon)).
\end{equation}
The error $\varepsilon$ is due to the approximation of a single $R_Z$ gate. The total error of the algorithm $\epsilon$ scales linearly with the number of gates that are decomposed approximately. Since the unitary evolution $U_w$ is decomposed into $4\log_2(N)-6$ $R_Y$ gates, for a quantum algorithm that uses $t$ time steps of the walk we have that the error per $R_Y$ gate is
\begin{equation}
\varepsilon = \frac{\epsilon}{(4\log_2(N)-6)t},
\end{equation}
so that the cost for a single $R_Y$ is expressed as
\begin{equation}
L_{R} = 4\log_2((4\log_2(N)-6)t/\epsilon)+O(\log(\log((4\log_2(N)-6)t/\epsilon)).
\end{equation}
If the algorithm is efficient then the number $t$ of time steps scales as a polynomial in $N$. An example is the quantum walk search algorithm with sinks, needing $\mathcal{O}(\sqrt{N/M})$ time steps for finding one of the $M$ marked nodes \cite{Portugal}. Therefore, the premise that $L_R$ scales slower than $\mathcal{O}(\log_2^3(N))$ is fulfilled.

\subsection{Bipartite graph}

A complete bipartite graph is shown in Figure \ref{F:bipartite_graph}. It is formed by two sets of nodes with $N_1$ and $N_2$ nodes, so that it has $N = N_1 + N_2$ nodes. Each node links to all the nodes of the other set, but there are no edges connecting two nodes inside the same set. Let us denote the set of indexes for the first $N_1$ nodes as $V_1$, and for the remaining $N_2$ nodes as $V_2$. The transition matrix in this case is:
\begin{equation}
G_{ji} =
\left\lbrace\begin{array}{c}
\displaystyle \ \ \ \ \frac{1}{N_2} \ \ \ \ \ \ \text{if} \ i \in V_1 \ \text{and} \ j \in V_2,\\
\\
\displaystyle \ \ \ \ \frac{1}{N_1} \ \ \ \ \ \ \text{if} \ i \in V_2 \ \text{and} \ j \in V_1,\\
\\
\displaystyle 0 \ \ \ \ \ \ \ \ \ \ \ \ \ \ \  \text{otherwise}.
\end{array}
\right.
\end{equation}

Let $n_1$ and $n_2$ be two integer numbers so that $N_1 = 2^{n_1}$ and $N_2 = 2^{n_2}$, with $n_2 \leq n_1$. In order to construct a quantum circuit for this graph we need $n = n_1+1$ qubits, so that the circuit simulates Szegedy quantum walk in an augmented graph with $2^n = 2N_1 \geq N_1+N_2$ nodes, but the original Hilbert space of the bipartite graph with $N_1+N_2$ nodes is recovered as an invariant subspace.\\

\begin{figure}[hbpt]
	\centering
	\includegraphics[scale=1]{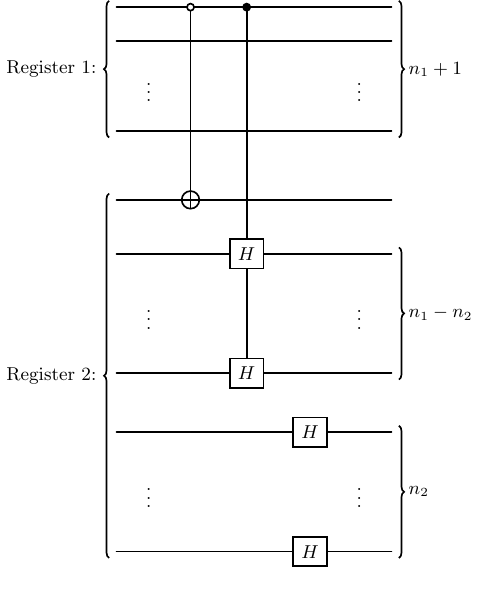}
	\caption{Quantum circuit for the update operator $\update$ for Szegedy quantum walk over a complete bipartite graph.}
	\label{F:update_bipartite}
\end{figure}

We have slightly modified the circuit found in the literature \cite{Q_circuits} in the case of the bipartite graph in order to obtain a single update operator $\update$ as per Figure \ref{F:szegedy_circuit_general}. Our quantum circuit for the update operator is shown in Figure \ref{F:update_bipartite}, and a proof of correctness is provided in Supplementary Material (SM) \cite{SM}. In this case the contribution to the count of $T/T^\dagger$ gates comes from the controlled array of $H$ gates. Since there is only a single control qubit, this controlled array corresponds to $n_1-n_2$ single-controlled-$H$ gates. Each of these provides $2$ $T/T^\dagger$ gates after rotating the $H$ gates into $X$ gates. Thus, the total number of $T/T^\dagger$ gates for the update operator $\update$ is:
\begin{equation}\label{L_K_bipartite}
\nT_{\update} = 2(n_1-n_2).
\end{equation}

Taking into account that the operator $D$ requires $n=n_1+1$ qubits, substituting in equation \eqref{T_count_general}, we obtain:
\begin{equation}
\nT_U = 14(n_1+1)-35 + 4(n_1-n_2) = 18n_1 - 4n_2 -21.
\end{equation}
Finally, we let the expression in terms of the number of nodes:
\begin{equation}
\nT_U = 18\log_2(N_1) - 4\log_2(N_2) -21,
\end{equation}
so it scales linearly with the logarithm of the number of nodes, and the homomorphic scheme can be implemented efficiently.

\section{Simulation results in Qiskit}\label{Simulation}

In this section, we show simulations of a quantum and a semiclassical Szegedy walk with a QHE scheme using IBM Qiskit simulator \cite{Qiskit}. Due to measurements of the Bell registers, the key-updating process is not deterministic, so that previous QHE implementations of Liang's scheme in Qiskit required precalculating all the possible mappings between the initial and the final key \cite{Pablo_Grover}, or was just done in a circuit with only Clifford gates \cite{QHE_grover}. For each $T/T^\dagger$ gate there are $2$ bits to consider, so that the number of possibilities to precalculate increased exponentially, making the simulation of an arbitrarily large circuit impossible. Moreover, an exponentially growing number of classically-controlled $S$ gates was required. In our case, thanks to the classical gates for the key-updating process described in Section \ref{QHE_circuits}, the encrypting key is updated during running time in Qiskit simulator, so that it is not necessary to precalculate anything and the running time of the simulation increases linearly.\\

\begin{figure}[hbpt]
    \centering
	\includegraphics[scale=0.4]{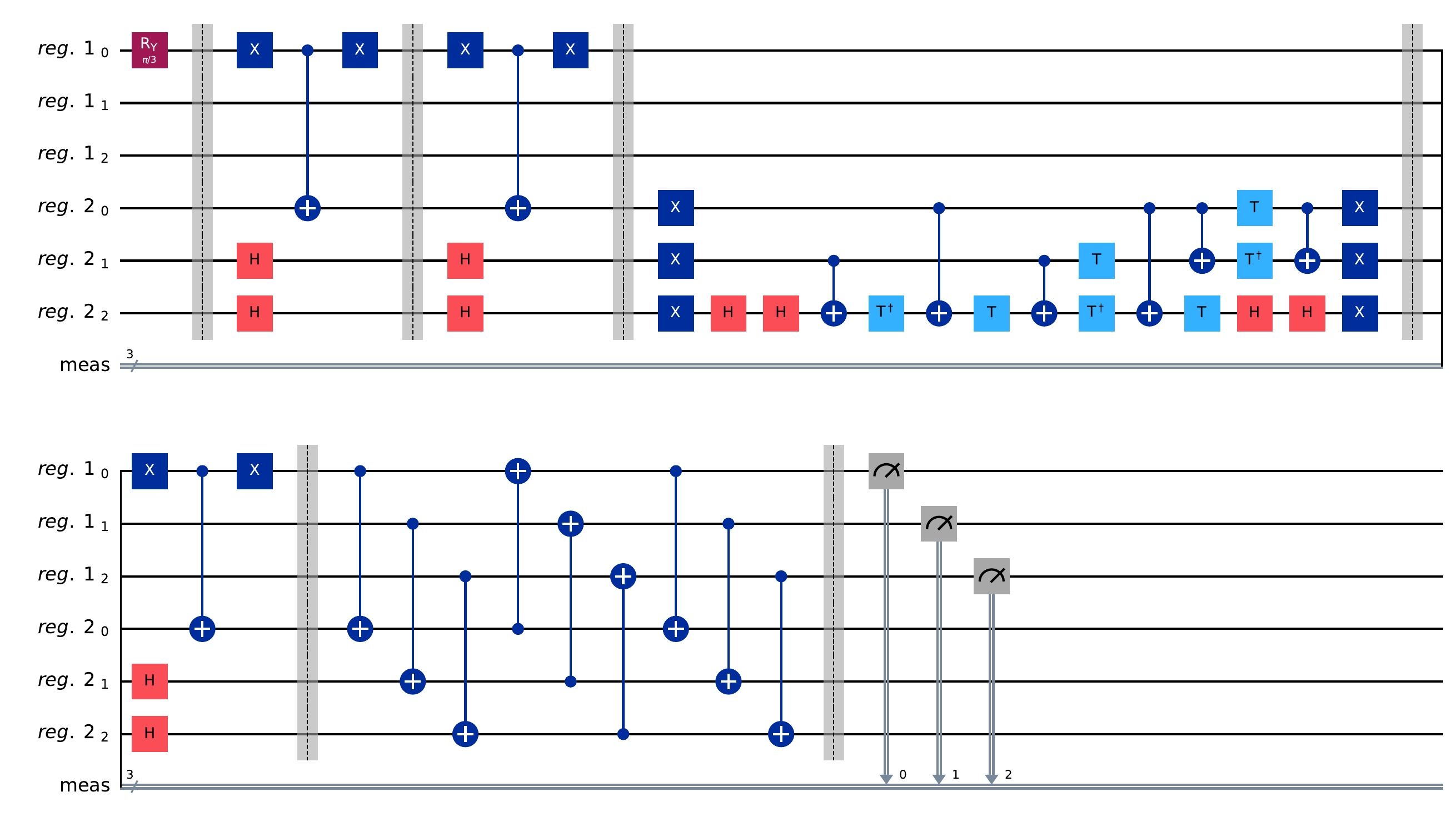}
	\caption{Quantum circuit for Szegedy quantum walk over a bipartite graph with $N_1=N_2=4$ nodes decomposed in Clifford+$T$ gates. Note that although Qiskit uses little-endian ordering, we construct the circuits using big-endian ordering, so that the resulting bitstring of the simulation must be reversed. The two first parts correspond to the $R_Y$ and $\update$ operators of Client. The rest of the circuit is the $U_w$ operator of Server. The last part of $U_w$ is the compilation of the three swap gates into $CNOT$ gates.}
	\label{F:naive_bipartite}
\end{figure}

First, we have simulated a time step of Szegedy quantum walk on the bipartite graph with $N_1 = N_2 = 4$, so that it is a graph with $N=8$ nodes. We choose this case because for $N_1=N_2$ the quantum circuit for the update operator $\update$ does not have any $T/T^\dagger$ gate, as shown in equation \eqref{L_K_bipartite}. Thus, since the circuit for this graph requires $n=3$ qubits per register, we only have the Toffoli gate of the diagonal operator $D$, so that the total number of $T/T^\dagger$ gates is $L=7$. This number is the minimum non-null number of $T/T^\dagger$ gates that we can have in a circuit, so this circuit is the easiest one to simulate. The decomposition of the algorithm in terms of Clifford+$T$ gates is shown in Figure \ref{F:naive_bipartite}. The initial state of the walk is chosen as $\left|\psi(0)\right> = \sqrt{0.75}\left|\psi_0\right> + \sqrt{0.25}\left|\psi_4\right>$. This state is created with an $R_Y$ gate that puts the first register in $\sqrt{0.75}\left|0\right>_1 + \sqrt{0.25}\left|4\right>_1$, and a posterior application of the update operator $\update$. Note that since Client does not perform QHE evaluation, the $R_Y$ gate does not need to be decomposed.\\

For the QHE implementation, in this case we need $14$ ancilla qubits, so that the total number of qubits required for the QHE implementation is $20$. Using the classical-quantum circuit blocks of Table \ref{tab:server-client} we have constructed the circuits for the three steps of the QHE scheme described in Section \ref{QHE}, in a similar manner as in Figure \ref{F:classical-quantum-circuits}. We have concatenated them in a single circuit so that it can be run in Qiskit simulator. At this point, we need to mention that Qiskit can implement classically-controlled gates, but it is not able to implement classical gates to classical bits in a straightforward manner. To overcome this issue, we have used two additional ancilla qubits. For each classical gate, using classically-controlled-$X$ gates we copy the value of classical bits into qubits starting in $\left|0\right>$, we apply the equivalent quantum gate between the qubits, and we measure them to copy the values back to the classical bits. Resetting these qubits after each classical gate, we can reuse them for the whole circuit. Further details for each classical gate, and also the random key initialization can be found in SM \cite{SM}. Also note that although qiskit uses little-endian ordering \cite{Qiskit}, we construct the circuits using big-endian ordering, so that the resulting bitstring of the simulation must be reversed.\\

In Table \ref{tab:bipartite_results} we show the results of seven repetitions of the simulation. We can observe how the initial key is updated into the final key at the end of the algorithm. Moreover, the last two rows highlighted in blue show how even in the case of starting with the same initial key, the final key is not obtained deterministically. In this case we obtain two different final keys, that depend on the measurement results of the Bell registers. In this algorithm we are obtaining the position of the walker measuring only the first register, so the classical results is a bitstring with only $3$ bits. Applying the element-wise XOR with the first three bits of the final $X$ key, we obtain the decrypted result.\\

\begin{table}[hbpt]
  \caption{Results for seven repetitions of the simulation of the QHE scheme applied to the quantum walk over the bipartite graph. We show the generated initial key, the results of measuring the Bell registers, the updated final key, and the results of measuring the qubits of the first register. These last results are decrypted using the first three bits of the final key associated to $X$ gates. The last two rows in blue show that the final key is obtained stochastically even if the initial key is the same.}
  \centering
    \begin{tabular}{cc|c|cc|cc}
    \hline
    \multicolumn{2}{c|}{Initial key} & \multirow{2}[4]{*}{$S^a$-rotated Bell measurements} & \multicolumn{2}{c|}{Final key} & \multicolumn{2}{c}{Result}\\
\cline{1-2}\cline{4-7}    $X$     & $Z$     &       & $X$     & $Z$     & Encrypted & Decrypted \\
    \hline
    $000|010$ & $011|100$ & $10|11|01|10|11|00|00$ & $010|000$ & $011|111$ & $101$ & $111$ \\
    $001|000$ & $011|110$ & $00|01|10|11|11|00|01$ & $010|001$ & $010|111$ & $110$ & $100$ \\
    $011|001$ & $011|000$ & $00|00|11|10|01|10|00$ & $100|011$ & $101|111$ & $000$ & $100$ \\
    $000|001$ & $001|101$ & $01|00|00|00|11|00|01$ & $001|000$ & $010|101$ & $101$ & $100$ \\
    $011|010$ & $010|101$ & $00|11|00|11|00|01|10$ & $000|011$ & $010|110$ & $101$ & $101$ \\
    \textcolor[rgb]{ .188,  .329,  .588}{$011|010$} & \textcolor[rgb]{ .188,  .329,  .588}{$111|011$} & \textcolor[rgb]{ .188,  .329,  .588}{$11|01|00|10|10|10|01$} & \textcolor[rgb]{ .188,  .329,  .588}{$111|011$} & \textcolor[rgb]{ .188,  .329,  .588}{$001|111$} & \textcolor[rgb]{ .188,  .329,  .588}{$011$} & \textcolor[rgb]{ .188,  .329,  .588}{$100$} \\
    \textcolor[rgb]{ .188,  .329,  .588}{$011|010$} & \textcolor[rgb]{ .188,  .329,  .588}{$111|011$} & \textcolor[rgb]{ .188,  .329,  .588}{$00|10|10|01|11|00|00$} & \textcolor[rgb]{ .188,  .329,  .588}{$010|011$} & \textcolor[rgb]{ .188,  .329,  .588}{$010|111$} & \textcolor[rgb]{ .188,  .329,  .588}{$110$} & \textcolor[rgb]{ .188,  .329,  .588}{$100$} \\
    \hline
    \end{tabular}
  \label{tab:bipartite_results}
\end{table}

\begin{figure}[hbpt]
    \centering
	\includegraphics[scale=0.75]{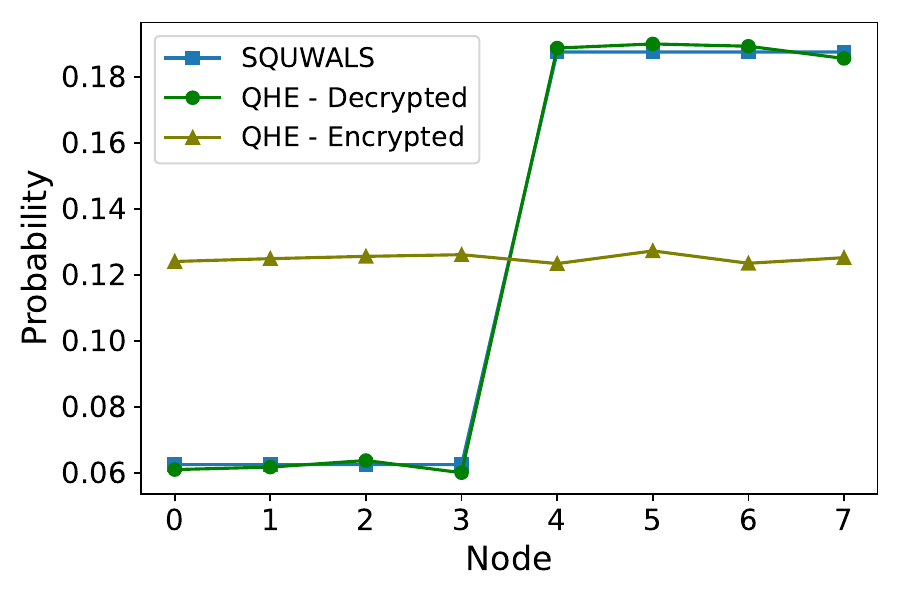}
	\caption{Probability distributions of the walker for the quantum walk over the bipartite graph using the QHE scheme and 20000 samples, before and after decrypting. The results are compared with the deterministic probability distribution obtained using SQUWALS \cite{Squwals}.}
	\label{F:bipartite_results}
\end{figure}

If we simulate the algorithm for a larger number of repetitions, we can sample the probability distribution of the walker. We have run the circuit for 20000 repetitions. The results are shown in Figure \ref{F:bipartite_results}. Before decrypting we obtain an homogeneous distribution. This is due not only to the randomness in the initial key generation, but also the randomness in the $S^a$-rotated Bell measurements. In order to evaluate the results after decrypting, we have simulated the walk algorithm deterministically using the python library SQUWALS, which provides an efficient classical simulator for Szegedy quantum walk \cite{Squwals}. We can see that both distributions are similar, and actually the differences are due to the stochasticity of Qiskit simulator for a finite number of repetitions. Thus, the implementation of the QHE scheme is successful.\\

In second place we simulate a semiclassical walk on a cycle graph with $N=8$ nodes. In this case the quantum circuit for the unitary evolution $U_w$ has $n=3$ qubits per register, so substituting in \eqref{L_U_cycle} it has $L_U = 77$ $T/T^\dagger$ gates. Thus, it would require $154$ ancilla qubits for the QHE scheme implementation of a single quantum step. For a classical simulator like Qiskit, whose memory requirements increase exponentially with the number of qubits, this is totally impossible. To overcome this issue, we propose a simplification of the simulation that requires only $2$ ancilla qubits for all the $T/T^\dagger$ gates, based on the simulation circuit implemented previously \cite{Pablo_Grover}. Due to the principle of deferred measurement, we showed that Client can wait to measure the Bell register after Server has finished. However, for the sake of simulation, Client could perform key-updating in parallel with the circuit of Server, and measure the Bell register at the moment Server performs the evaluation of the $T/T^\dagger$ gate. Thus, we can reset the Bell qubits after measurement and reuse them for all the $T/T^\dagger$ gates. Further details about this simplification, and results showing that both simulation implementations provide the same results are provided in SM \cite{SM}.\\

\begin{figure}[hbpt]
	\centering
	\subfigure[]{\includegraphics[scale=0.55]{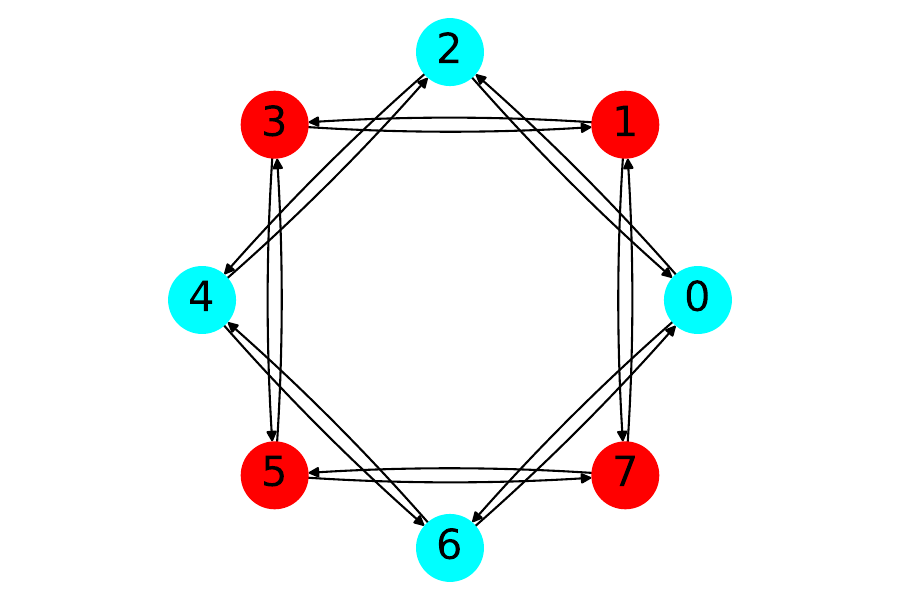}\label{F:semiclassical_graph}}
	\subfigure[]{\includegraphics[scale=0.55]{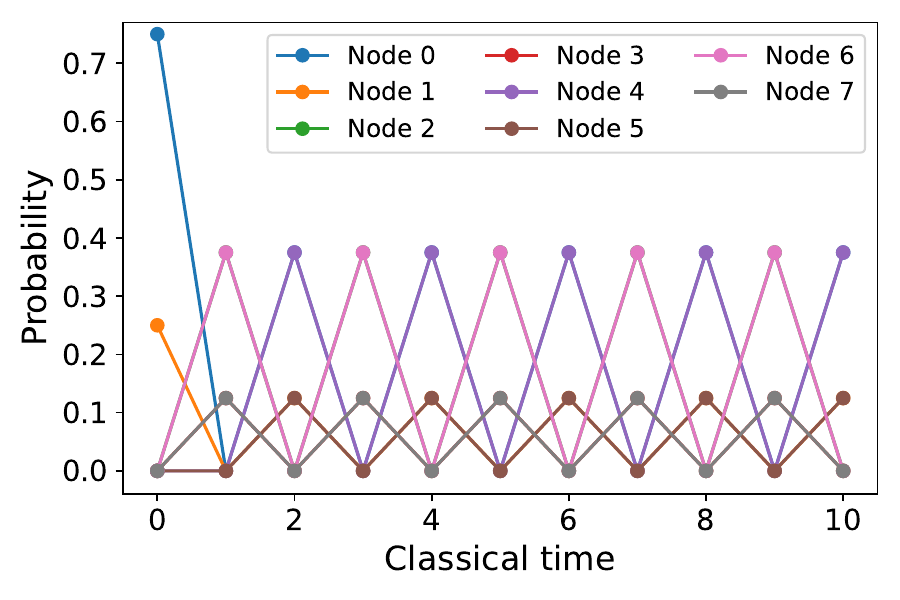}\label{F:semiclassical_squwals}}
	\subfigure[]{\includegraphics[scale=0.55]{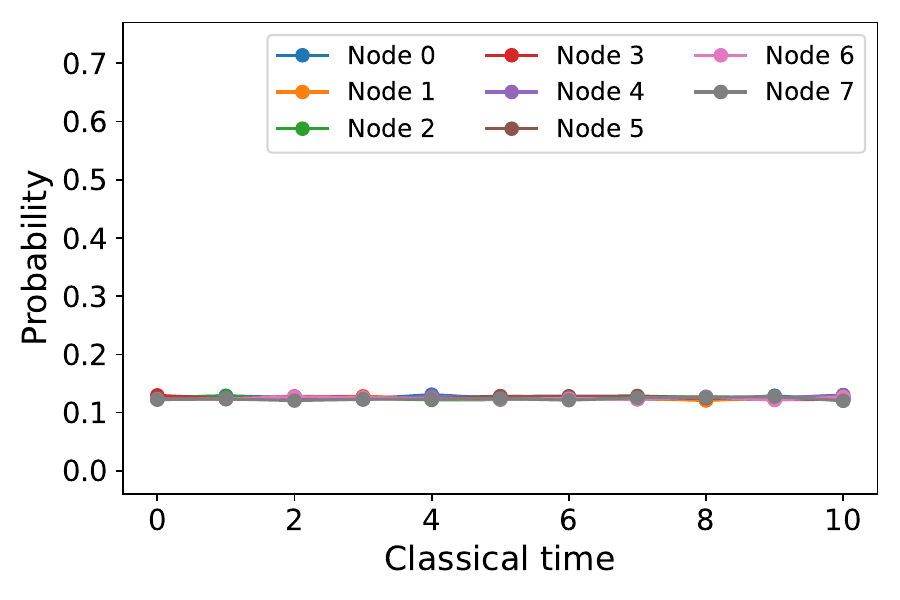}\label{F:semiclassical_results_1}}
    \subfigure[]{\includegraphics[scale=0.55]{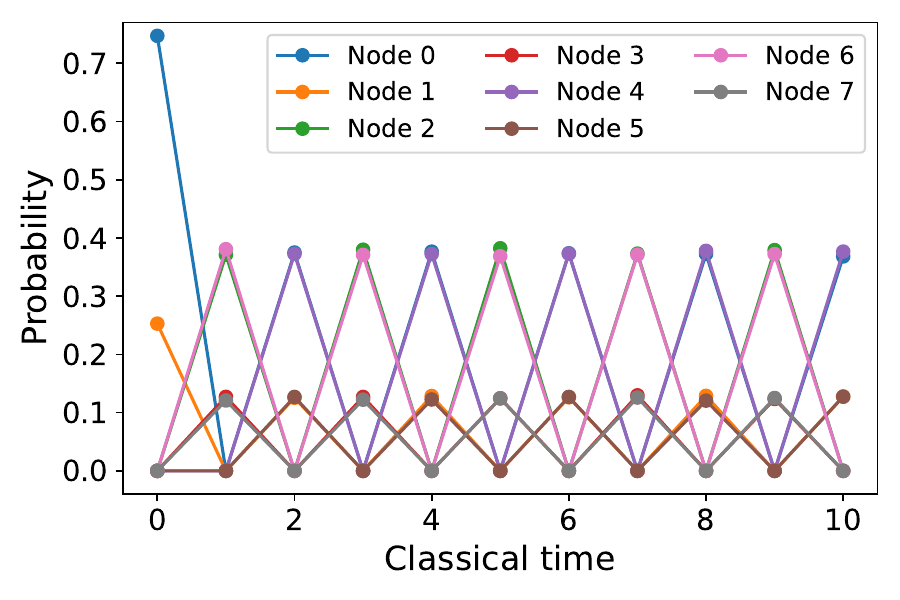}\label{F:semiclassical_results_2}}
	\caption{Semiclassical graph for the semiclassical walk over the cycle with $N=8$ nodes for $t_q=2$. The graph has been plotted using the python library NetworkX \cite{NetworkX}. b) Probability of the walker being at each of the eight nodes at each classical time step of the semiclassical walk in a) obtained deterministically using SQUWALS \cite{Squwals}. c)-d) Probability of the walker being at each of the eight nodes at each classical time step of the semiclassical walk in a) sampled with 20000 repetitions of the QHE scheme simulation before decrypting and after decrypting respectively.}
	\label{F:N=6}
\end{figure}

For the semiclassical walk on the cycle, we choose a quantum time $t_q = 2$. In this case, the semiclassical walk is equivalent to a classical walk over two independent square graphs \cite{Semiclassical}, whose representation is shown in Figure \ref{F:semiclassical_graph}. As the initial classical probability vector we take $(0.75,0.25,0,0,0,0,0,0)^T$, whose representing quantum state is created by Client with a $R_Y$ gate. We have simulated the semiclassical walk for $t_c = 10$ classical steps. The results of the deterministic simulation using SQUWALS \cite{Squwals} are shown in Figure \ref{F:semiclassical_squwals}. We show the probability of the walker being at each node for each classical step. Since both squares have an even number of nodes, the walker alternates between odd and even nodes at each time step. Moreover, since both squares are not connected, the total probability inside them is conserved. The results of the QHE scheme implementation using Qiskit for 20000 repetitions are shown in Figure \ref{F:semiclassical_results_1} before decrypting, and in Figure \ref{F:semiclassical_results_2} after decrypting. As expected, before decrypting we obtain an homogeneous distribution, whereas after decrypting we obtain similar results to the deterministic simulation. Thus, the QHE scheme simulation is successful also for the semiclassical walks, which include reset and measurement operations. The fact that the QHE scheme works perfectly for both types of walks demonstrates the correctness of the scheme. We want to highlight that these simulations constitute the first simulations of Liang's scheme where the number of classically controlled $S$ gates is linear and also where the keys are updated as the simulation is run. Furthermore, the Python library CQC-QHE constructed from the previous simulations represents an important contribution to the field of QHE simulation, since it can be applied to a plethora of quantum circuits. Therefore, QHE simulations of many existing quantum algorithms can easily be tested, which could represent an important leap in the field.\\

\section{Conclusions}\label{Conclusions}

In this work, Liang's quasi-compact QHE scheme \cite{Liang} has been reinterpreted in terms of classical-quantum circuits. The resulting scheme is still perfectly secure, non-interactive, $\mathcal{F}$-homomorphic and quasi-compact. The complexity of the decryption procedure depends on $L$, the number of $T/T^{\dagger}$ gates. Moreover, we have extended the analysis to prove that it can also be applied to semiclassical circuits, using the key-updating rules for the homomorphic evaluation of the reset and intermediate measurement operations.\\

We have studied the efficiency of this QHE scheme applied to both quantum and semiclassical walks based on Szegedy's quantization on certain types of networks: cyclic graphs, complete graphs and bipartite graphs. As it has been shown, the value of $L$ for the different graphs studied grows slower than any linear function. In particular, $L$ scales as $\mathcal{O}(\log_2^3(N))$ for the cyclic graph and the complete graph, and it scales as $\mathcal{O}(\log_2(N))$ for the bipartite graphs. The scheme is only efficient for circuits with a polynomial number of $T/T^{\dagger}$ gates. Therefore, the Szegedy quantum walks studied are perfect examples of algorithms that can be evaluated homomorphically with perfect security and non-interaction in an efficient manner.\\

Recently a graph-phased Szegedy quantum walk has been developed, including complex-phase extensions \cite{Graph-phased}. In this case the diagonal operator $D$ is substituted by several multi-controlled phase gates. Since the phase gates can be converted into $R_Z$ gates up to a global phase, these can also be decomposed efficiently into Clifford+$T$ gates. Therefore, an efficient implementation of the QHE scheme should also be possible in this new paradigm.\\

Using this QHE scheme, first we have simulated in Qiskit a simple quantum circuit that implements the walk on a bipartite graph homomorphically to check that the scheme works. We showed how to apply classical-quantum gates in the Qiskit simulation in order to correct the phase errors and update the keys as the gates were applied. To check that the obtained results are correct the python library SQUWALS was used to simulate the walk algorithm deterministically. The correct results can always be successfully decrypted, independently of the initial key used in encryption. After that, we have simulated a quite more complicated circuit for a semiclassical walk on a cycle graph. This simulation contains intermediate measurement and reset operations. Once again, the walk was simulated deterministically using SQUWALS to check the correctness of the decrypted results of the QHE simulation. These positive results demonstrate the correct functioning of the QHE protocol, even in the presence of non-unitary operations.\\ 

These simulations represent an improvement compared to previous simulations of the QHE scheme, since those simulations were limited to Clifford gates or had an exponential number of classically controlled $S$ gates, requiring the calculation of the final keys beforehand. Our simulations make use of classical gates to simulate the key-updating functions at running time. This is done for all gates, so the decryption procedure scales linearly with the total number of gates rather than $L$. The reason for this is that Qiskit does not allow arbitrary functions over classical bits. This method of QHE simulation is a proof of concept and we leave the compilation of the Clifford composite functions in XOR gates as future work.\\ 

We have used our results to create a Python library based on Qiskit, named CQC-QHE, able to construct and classically simulate the classical-quantum circuits required for quantum homomorphic encryption. Currently user must provide the circuit compiled into Clifford+$T$ gates, or with multi-controlled-$X$ gates with black dots. The automatic decomposition of circuits containing other types of gates, such as controlled-$H$ gates or $R_Y$ gates, is another research line that will be expanded as future work.\\

Finally, we want to mention that more optimizations can be applied regarding the decomposition of controlled gates, such as decomposing the multi-controlled-X gate using the relative phase Toffoli gate \cite{Relative_Toff} or the recent multi-controlled quantum gates implementations \cite{Multi_controlled_decomposition}. These implementations would result in a reduced number of $T/T^{\dagger}$ gates and ancilla qubits. In any case, since the implementations we have constructed are already efficient for the Szegedy graphs studied, we leave the study of further optimizations as future work.\\

So far all the graphs we have studied are actually toy models with limited applications, as for example sink nodes search on the complete graph \cite{Portugal}. In the future we need to analyze the QHE scheme applied on more complex quantum walks with real applications, as for example those used for optimizations problems \cite{Qfold,QMS,GWQMA,qBIRD}. Furthermore, our modified QHE scheme can be applied not only to quantum walks, but to a wide variety of quantum algorithms, so we expect to find soon further use case applications.\\

\section{Data Availability Statement}

The python library CQC-QHE is available on GitHub: \url{https://github.com/OrtegaSA/CQC-QHE-repo}.\\

There is a guide about how the different functions work, and tutorials implementing the quantum and semiclassical walks shown in this paper.

\section*{Acknowledgments}

S.A.O. and P.F. contributed equally to this work. We acknowledge the support from the Spanish MINECO grants MINECO/FEDER Projects, PID2021-122547NB-I00 FIS2021, the “MADQuantum-CM” project funded by Comunidad de Madrid and by the Recovery, Transformation, and Resilience Plan – Funded by the European Union - NextGenerationEU and Ministry of Economic Affairs Quantum ENIA project. This work has also been financially supported by the Ministry for Digital Transformation and of Civil Service of the Spanish Government through the QUANTUM ENIA project call – Quantum Spain project, and by the European Union through the Recovery, Transformation and Resilience Plan – NextGenerationEU within the framework of the Digital Spain 2026 Agenda. M. A. M.-D. has been partially supported by the U.S.Army Research Office through Grant No. W911NF-14-1-0103. S.A.O. acknowledges support from Universidad Complutense de Madrid - Banco Santander through Grant No. CT58/21-CT59/21. P.F. acknowledges support from a MICINN contract PRE2019-090517
(MICINN/AEI/FSE,UE).

\bibliography{MiBiblio}

\begin{thebibliography}{10}
\bibitem{QT}
{By quantum time we refer to a parameter that determines the duration of the
  quantum evolution, rather than a quantum operator.}
\end{thebibliography}
\bibliographystyle{unsrt}

\begin{appendix}

\section{Clifford+$T$ gates}\label{A:gates}

The usual Clifford gates used in this work are $\{X,Z,H,S,S^\dagger,CNOT\}$ \cite{Nielsen,MA_review_modern_physics}. Gates $X$ and $Z$ are the single-qubit Pauli gates. The Hadamard gate is $H=\frac{X+Z}{\sqrt{2}}$. Their matrix representation is given by:
\begin{equation}
X=\begin{pmatrix}
0& 1 \\
1 & 0
\end{pmatrix}, \quad Z=\begin{pmatrix}
1& 0 \\
0 & -1
\end{pmatrix}, \quad
H=\frac{1}{\sqrt{2}}\begin{pmatrix}
1& 1 \\
1 & -1
\end{pmatrix}.
\label{matrix1}
\end{equation}
The phase gate $S=\sqrt{Z}$, its conjugate $S^\dagger$, and $CNOT$ gate matrices are: 
\begin{equation}
S=\begin{pmatrix}
1& 0 \\
0 & i
\end{pmatrix},
\quad S^\dagger=\begin{pmatrix}
1& 0 \\
0 & -i
\end{pmatrix},
\quad CNOT=\begin{pmatrix}
1& 0&0&0 \\
0& 1&0&0 \\
0& 0&0&1 \\
0& 0&1&0 \\
\end{pmatrix}.
\end{equation}
In order to perform universal quantum computation, a non-Clifford gate must be added to the Clifford gates \cite{Nielsen}. The matrix representations of the $T$ gate and its conjugate $T^\dagger$ are:
\begin{equation}
T= \begin{pmatrix}
1& 0 \\
0 & e^{i\frac{\pi}{4}}
\end{pmatrix}, \quad
T^{\dagger}= \begin{pmatrix}
1& 0 \\
0 & e^{-i\frac{\pi}{4}}
\end{pmatrix}.
\end{equation}

\section{Gate teleportation}\label{A:gate_teleportation}

An EPR state is an entangled quantum state given by $\ket{\Phi_{00}}=\frac{1}{\sqrt{2}} (\ket{00}+\ket{11})$ \cite{Einstein}. The state $\ket{\Phi_{00}}$ can be constructed by using a $H$ gate followed by a $CNOT$ acting on the state $\ket{00}$.	
From this entangled state, we can express the well known four Bell states in a compact expression:
\begin{equation}
\ket{\Phi_{ba}}=(Z^bX^a \otimes I)\ket{\Phi_{00}}, \forall a, b \in {0,1}.
\end{equation}

Quantum teleportation \cite{Teleportation} is a technique that allows the transferring of quantum states between a sender and a receiver using a quantum communication channel. In this procedure, Alice wants to send Bob a qubit in the state $\ket{\psi}$, so they first share an entangled pair of qubits in the state $\ket{\Phi_{00}}$. This way both have a qubit from the EPR pair. Then Alice performs a quantum measurement in the Bell basis using the two qubits she has, $\ket{\psi}$ and one qubit of the pair $\ket{\Phi_{00}}$. Due to entanglement, Bob can now obtain the original state $\ket{\psi}$ if he applies the correct sequence of quantum gates (a combination of $X$ and $Z$) to the qubit in his possession.\\

For any single-qubit gate $U$, the ``$U$-rotated Bell basis'' can be defined as \cite{U_rotated}: $\Phi(U)=\{\ket{\Phi(U)_{ba}}, a,b \in \{0,1\} \}$, where $\ket{\Phi(U)_{ba}}=(U^{\dagger} \otimes I)\ket{\Phi_{ba}}=(U^{\dagger}Z^bX^a \otimes I)\ket{\Phi_{00}}$.\\

\begin{figure}[]
	\includegraphics[width=0.48\textwidth]{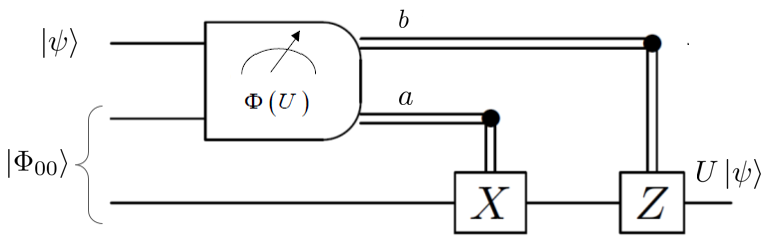}
	\centering
	\caption{Quantum circuit for gate teleportation. The box represents the quantum measurement Alice performs on the qubit $\ket{\psi}$ and one qubit of the pair $\ket{\Phi_{00}}$. The measurement basis chosen is the $U$-rotated Bell basis $\Phi(U)$. Depending on the measurement results, $a$ and $b$, Bob applies $X^{a}$ and $Z^{b}$ in order to obtain $U \ket{\psi}$.}
	\label{gatetelepory}
\end{figure}

For a single qubit we have the following expression for quantum teleportation:
\begin{equation}
\ket{\psi} \otimes \ket{\Phi_{00}} = \sum_{a,b\in \{0,1\}} \ket{\Phi_{ba}} \otimes X^aZ^b \ket{\psi}.
\end{equation}
It can easily be extended for the ``$U$-rotated Bell basis'':
\begin{equation}
\ket{\psi} \otimes \ket{\Phi_{00}} = \sum_{a,b\in \{0,1\}} \ket{\Phi(U)_{ba}} \otimes X^aZ^bU \ket{\psi},
\label{gate teleport}
\end{equation}
where $U$ is any single-qubit gate. Then equation $\eqref{gate teleport}$ describes gate teleportation. Just like in the usual quantum teleportation, Alice and Bob first share an entangled EPR pair in the state $\ket{\Phi_{00}}$. Then Alice prepares a qubit in the state $\ket{\psi}$ and performs a ``$U$-rotated Bell measurement''. This is simply a quantum measurement in which the $U$-rotated Bell basis is selected as its measurement basis. She performs the measurement on the two qubits she possesses, one in the state $\ket{\psi}$ and one of the EPR pair in the state $\ket{\Phi_{00}}$. From this measurement she obtains the results $a$ and $b$. After the measurement Bob's qubit transforms into the state $X^aZ^bU \ket{\psi}$. Next, Alice tells Bob the results of her measurement, so Bob can apply the correct Pauli $X$ and $Z$ operators to finally obtain $U \ket{\psi}$. The gate teleportation procedure is represented in figure \ref{gatetelepory}.\\

\end{appendix}

\clearpage

\onecolumngrid
\newpage
\begin{center}
	\textbf{\large Supplementary Material: Implementing Semiclassical Szegedy Walks in Classical- Quantum Circuits for Homomorphic Encryption}
\end{center}
\setcounter{equation}{0}
\setcounter{section}{0}
\setcounter{figure}{0}
\setcounter{table}{0}
\setcounter{page}{1}
\makeatletter
\renewcommand{\theequation}{S\arabic{equation}}
\renewcommand{\thesection}{S\Roman{section}}
\renewcommand{\thefigure}{S\arabic{figure}}
\renewcommand{\bibnumfmt}[1]{[S#1]}
\renewcommand{\citenumfont}[1]{S#1}

\begin{center}
Sergio A. Ortega, Pablo Fernández and Miguel A. Martin-Delgado
\end{center}

\newcommand{\updateSM}{V}
\newcommand{\vertexset}{\mathcal{V}}

\section{Classical gates in Qiskit}

\begin{figure}[H]
	\centering
	\includegraphics[scale=0.52]{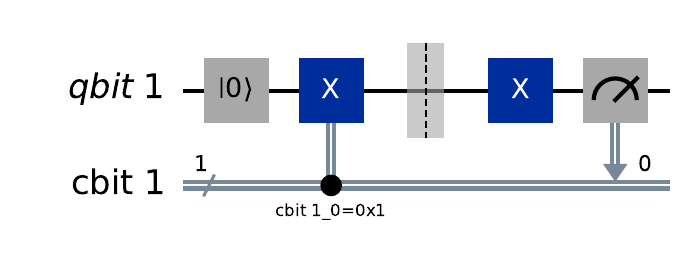}
	\caption{Quantum circuit for the emulation of a classical $NOT$ gate using an ancilla qubit.}
	\label{F:classical_X}
\end{figure}

Qiskit allows quantum gates that are controlled by classical bits. However, it currently does not allow the application of classical gates over classical bits. To overcome this issue, we use ancilla qubits to emulate these classical gates. These ancilla qubits can be loaded with the value of the classical bits, we can apply the equivalent quantum gate to the desired classical gate, and finally measure them to copy the results back to the bits. Let us start with a classical bit-flip or $NOT$ gate. The classical-quantum circuit emulating this gate with an ancilla qubit is shown in Figure \ref{F:classical_X}. We reset the ancilla qubit so that it is in the state $\left|0\right>$, and we apply a $X$ gate controlled by the classical bit. Thus, if the bit has the value $a \in \lbrace{0,1\rbrace}$, qubit takes the state $\left|a\right>$. The quantum gate equivalent to the classical not is the $X$ gate. We apply it to the qubit, and finally measure its value back to the bit. Note that since the qubit was previously reset to $\left|0\right>$, it is always either on $\left|0\right>$ or $\left|1\right>$ after the application of the $X$ gates, so that the measurement is deterministic and the emulation of the classical gate is correct.

\begin{figure}[H]
	\centering
	\subfigure[]{\includegraphics[scale=0.52]{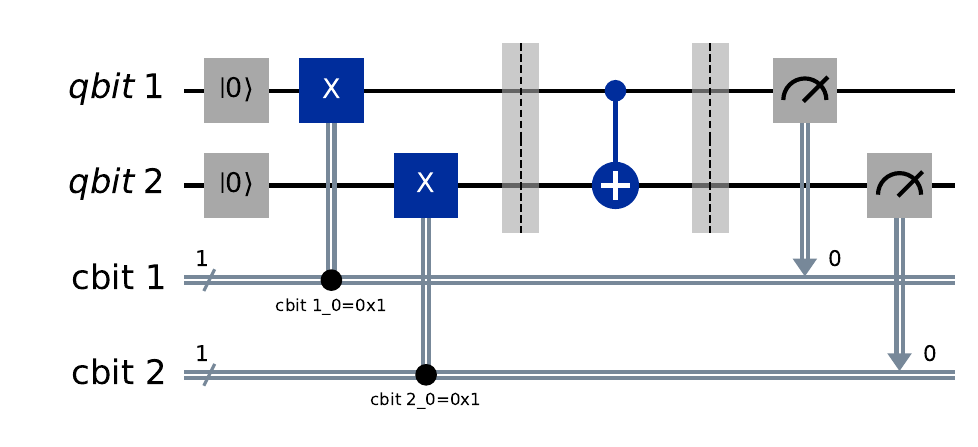}}
	\subfigure[]{\includegraphics[scale=0.52]{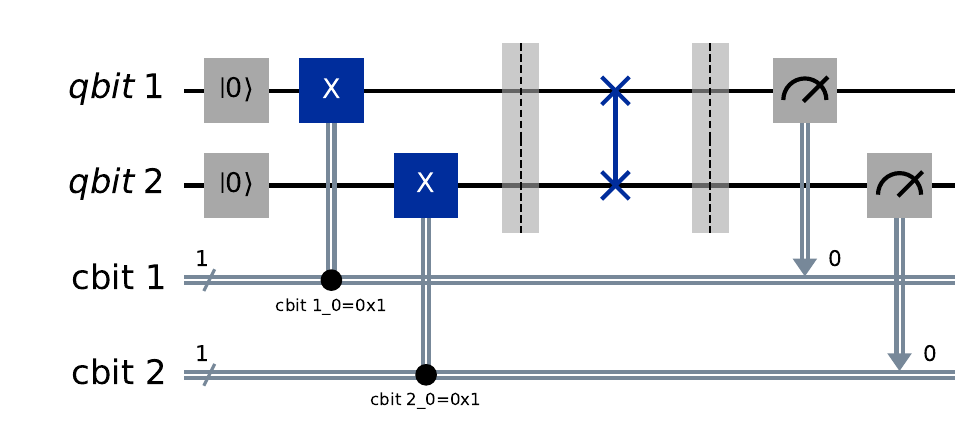}}
	\caption{a) Quantum circuit for the emulation of a classical $CNOT$ gate using two ancilla qubits. b) Quantum circuit for the emulation of a classical swap gate using two ancilla qubits.}
	\label{F:classical_two_bits}
\end{figure}

For classical gates acting over two bits, as the classical $CNOT$ or the classical swap gates, we need two ancilla qubits. The classical-quantum circuits emulating these gates are shown in Figure \ref{F:classical_two_bits}. Again, we reset the ancilla qubits, and copy the values of the classical bits using classically-controlled-$X$ gates. After that, we apply the equivalent quantum gate, a $CNOT$ or a swap gate, to the ancilla qubits, and measure them back to the classical bits.

\begin{figure}[H]
	\centering
	\includegraphics[scale=0.52]{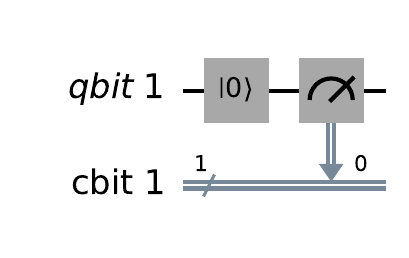}
	\caption{Quantum circuit for the emulation of a classical reset using an ancilla qubit.}
	\label{F:classical_reset}
\end{figure}

Another classical gate we are interested in is the reset gate, that erases the value of a classical bit letting it in $0$. In this case, we reset an ancilla qubit to $\left|0\right>$ and measure it to the classical bit. The classical-quantum emulation circuit is shown in Figure \ref{F:classical_reset}.

\begin{figure}[H]
	\centering
	\includegraphics[scale=0.52]{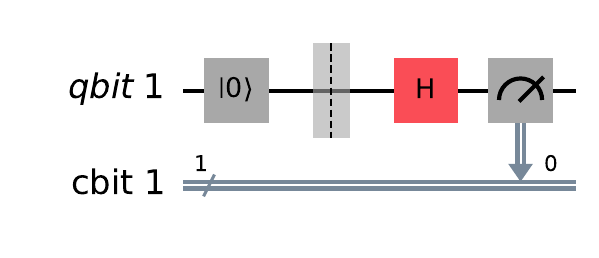}
	\caption{Quantum circuit for the emulation of a classical random initialization using an ancilla qubit.}
	\label{F:classical_H}
\end{figure}

Finally, we also want to be able to initialize the classical bits at $0$ or $1$ at random in order to create an initial secret key at running time. In this case the quantum analogue gate is the Hadamard $H$ gate. Although this gate is purely quantum and the qubit ends up in a coherent superposition, if we measure it the classical bit obtains either $0$ or $1$ at random with a probability of $50\%$ for each value. The classical-quantum emulation circuit is shown in Figure \ref{F:classical_H}.

Since we need to reset the ancilla qubits before each classical gate emulation, and these ancilla qubits play no more role after the measurement, we can reuse them for all the classical gates in the circuit. Thus, in total we need only two ancilla qubits. Moreover, we can even avoid the addition of such qubits in the simulation. As we explained in the main text, in our QHE scheme Client measures the quantum state in the computational basis and then decrypts the classical result. This measurement can be performed before all the classical gates used for the key-updating process. Since after measurement the main qubits of the circuit play no role, we can reset them and use two of them as the ancilla qubits for the classical gates. Thus, we save memory and time resources by a factor of four, since we do not add two new qubits to the system. Of course, these simplification can only be done provided that the main circuit has more than a single qubit, which is the most common situation.

\section{Simplified simulation}

As mentioned in the main paper, the memory requirements of a classical simulator like Qiskit increase exponentially with the number of qubits. Thus, our simulation method for the QHE scheme described in this work, which uses two ancilla qubits for each $T/T^\dagger$ gate is not suitable for an arbitrarily large number of $T/T^\dagger$ gates.\\

\begin{figure}[H]
	\centering
	\includegraphics[scale=1]{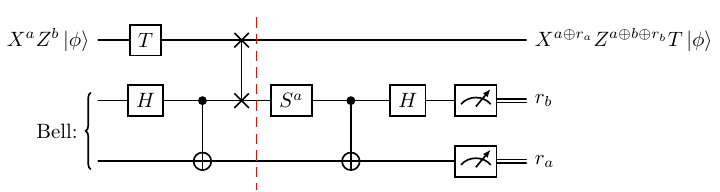}
	\caption{Homomorphic evaluation scheme for a $T$ gate using gate teleportation. The part at the left of the red line is performed by Server, and the part at the right by Client once Server has finished running the quantum algorithm.}
	\label{F:T-evaluation_SM}
\end{figure}

In Figure \ref{F:T-evaluation_SM} we show again the homomorphic evaluation scheme for a $T$ gate. We said that due to the principle of deferred measurement, Client could wait to measure the Bell register after Server has finished. Thus, the part on the right of the red line would be implemented by Client once Server has finished. However, for the sake of simulation, the part of Client can be implemented right after the part of Server. To do so, we need to know at that moment of the simulation the value of classical bit $a$, stored in the classical bit $x$ associated with the qubit. Thus, the key-updating process prior to that $T$ gate must have been performed too.

Then, for each gate of the quantum algorithm, we need to implement its quantum evaluation scheme circuit in parallel with the associated classical-quantum circuit for key-updating, which were shown in Tables I and II. After the measurement of a quantum Bell register, these qubits play no more role in the simulation. Thus, they can be reset and reused for the homomorphic evaluation of the next $T/T^\dagger$ gate. Thus, with only two ancilla qubits we can represent all the Bell registers that would be needed in a real implementation of the QHE scheme. Therefore, the memory requirements do not depend on the number of $T/T^\dagger$ gates, and we can simulate the scheme for any circuit as long as the number of qubits of the original algorithm allows it.\\

\begin{figure}[H]
	\centering
	\includegraphics[scale=1]{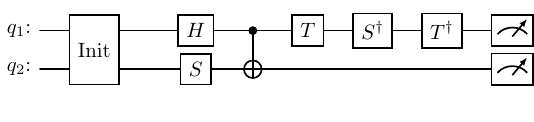}
	\caption{Example of a quantum algorithm with $2$ $T/T^\dagger$ gates and Clifford gates applied over two qubits.}
	\label{F:naive_circuit_SM}
\end{figure}

Let us see an example using the same example algorithm of the main text, whose quantum circuit we show again in Figure \ref{F:naive_circuit_SM}.

\begin{figure}[H]
	\centering
    \makebox[10pt][c]{
	\includegraphics[scale=1]{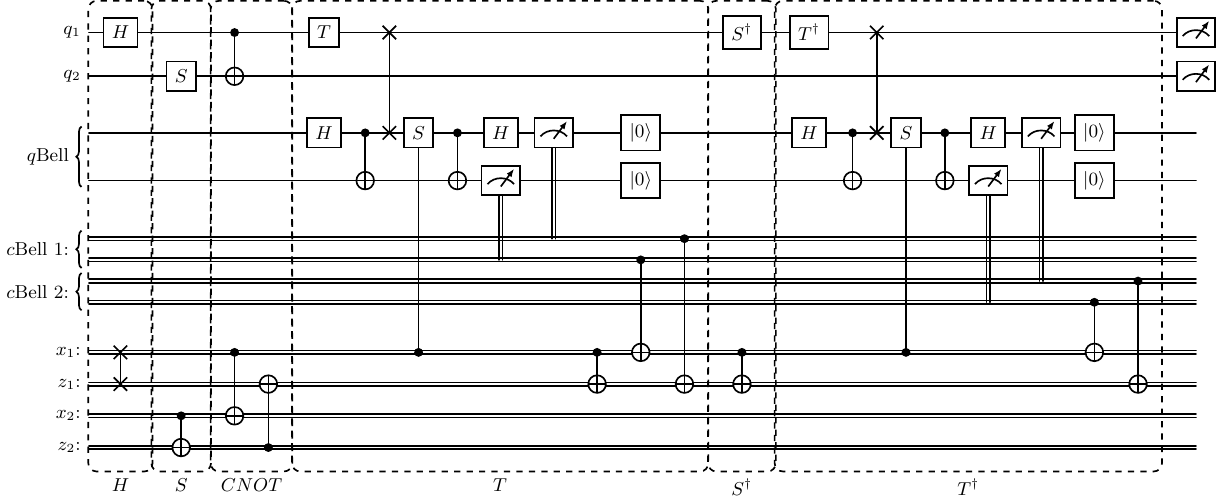}
    }
	\caption{Classical-quantum circuit for the simplified simulation of steps 2 and 3 of the quantum homomorphic encryption scheme applied over the example in Figure \ref{F:naive_circuit_SM}. The evaluation schemes of Sever are applied in parallel with the key-updating functions of Client. The qubits of the register $q$Bell start in state $\left|0\right>$. Quantum bits are represented by single lines, whereas classical bits are represented by double lines.}
	\label{F:server_client_circuit}
\end{figure}

The classical-quantum circuit of Client for the initialization of the algorithm in Step 1 would be the same that the one we described in Figure 3(a). However, the circuits of Server and Client in Steps 2 and 3, shown in Figures 3(b) and 3(c), are combined into a single one that performs the key-updating subroutines in parallel with the quantum algorithm. This classical-quantum circuit is shown in Figure \ref{F:server_client_circuit}. As we can see, for each quantum gate in the original circuit we implement its homomorphic evaluation scheme along with its key-updating function, using the same two ancilla qubits for all the $T/T^\dagger$ gates.

Note that for this simulation procedure the main qubits are measured at the end of the circuit. Thus, we cannot take two of them as ancilla qubits for the classical gates emulation, as explained in the previous section. However, since the quantum Bell register is reset after each measurement, we can also use its two qubits for all the classical gates. Thus, for the simplified simulation we can also avoid adding two additional qubits to the system.\\

\begin{figure}[H]
	\centering
	\subfigure[]{\includegraphics[scale=0.55]{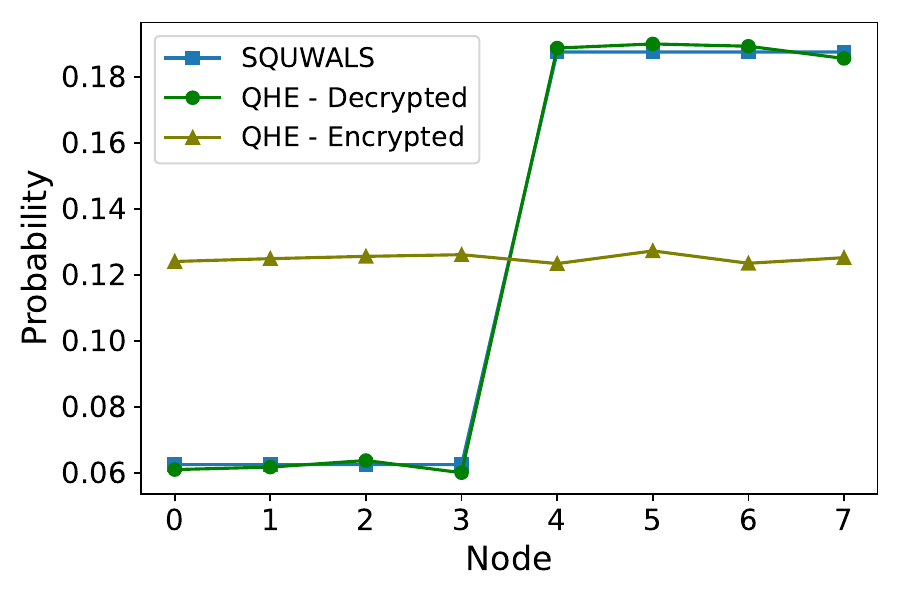}\label{F:bipartite_results_full}}
    \subfigure[]{\includegraphics[scale=0.55]{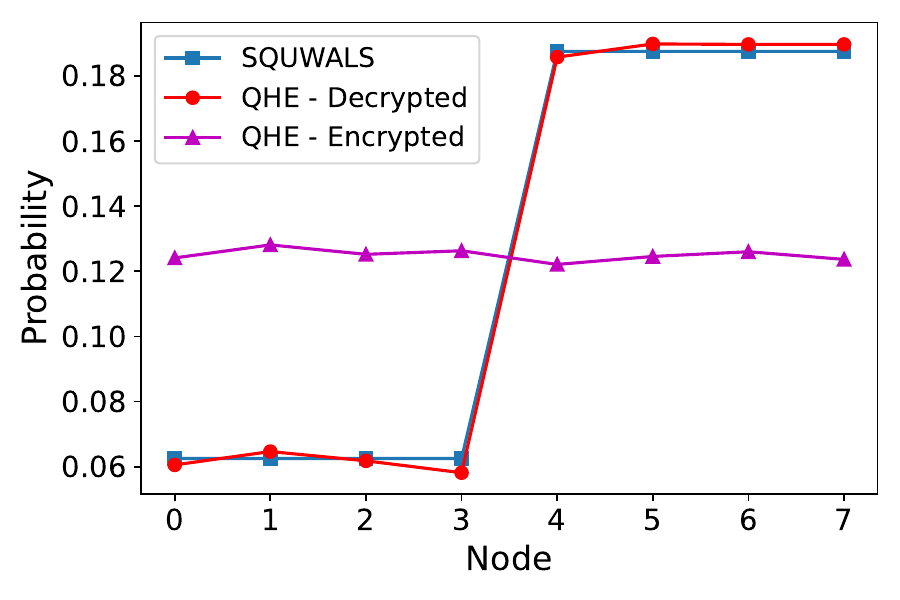}\label{F:bipartite_results_simplified}}
	\caption{Probability distributions of the walker for the quantum walk over the bipartite graph using the QHE scheme and 20000 samples, using a) the realistic simulation and b) the simplified simulation. The results are compared with the deterministic probability distribution obtained using SQUWALS.}
	\label{F:bipartite_results_SM}
\end{figure}

To check that the simplified simulation provides the same result than the simulation with all the ancilla qubits, we compare the results obtained with both methods for the simulation of the bipartite graph of Section V in Figure \ref{F:bipartite_results_SM}. The results in both cases are similar, and the differences are due again to the fact that we are sampling the probability distributions with a finite number of repetitions.

\begin{figure}[H]
    \centering
	\includegraphics[scale=0.5]{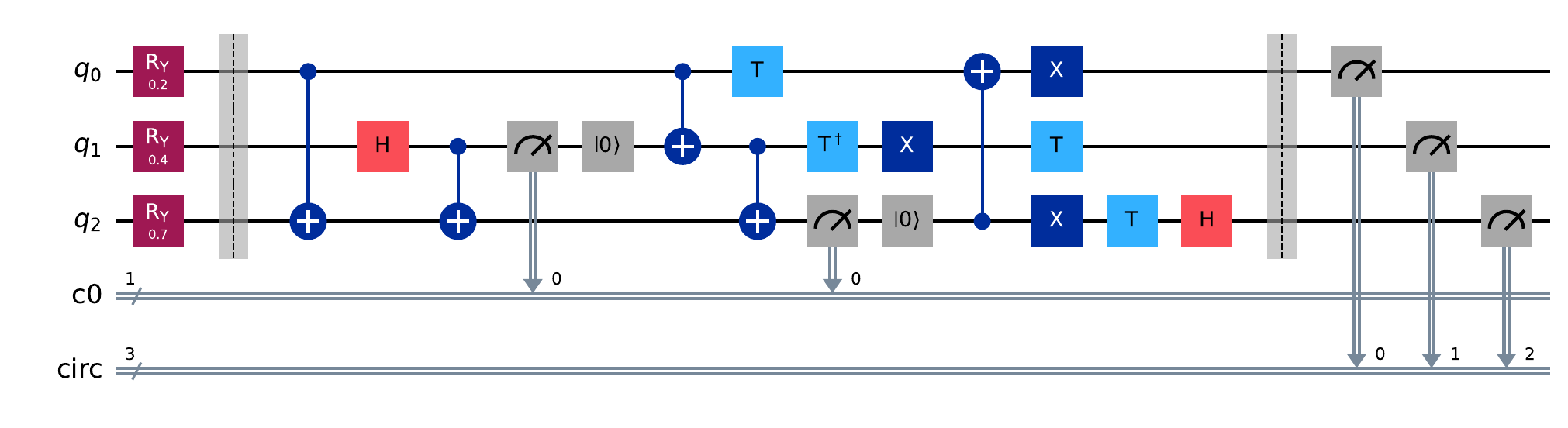}
	\caption{Example circuit with measurement and reset operations. The part of Client corresponds to the $R_Y$ gates before the first barrier. Note that although Qiskit uses little-endian ordering, we construct the circuits using big-endian ordering, so that the resulting bitstring of the simulation must be reversed.}
	\label{F:naive_example}
\end{figure}

Finally, we also check that the simplification works for an artificial circuit which also contains measurement and reset operations. The quantum circuit is shown in Figure \ref{F:naive_example}.

\begin{figure}[H]
	\centering
	\includegraphics[scale=0.7]{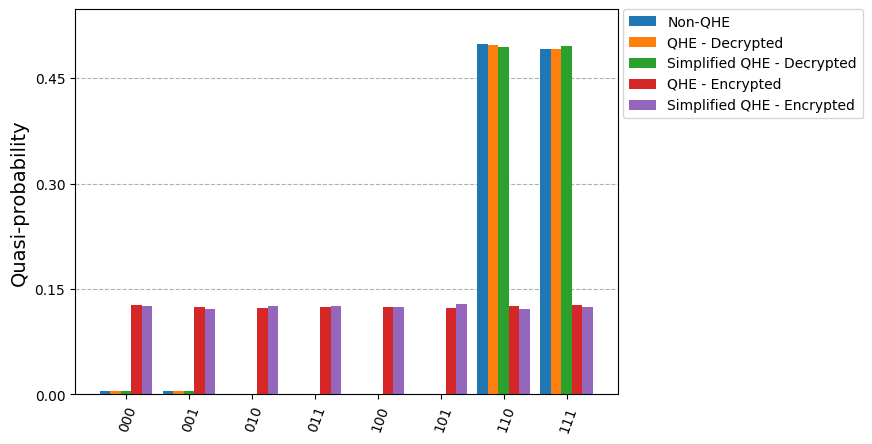}
	\caption{Results of the QHE scheme of the quantum circuit in Figure \ref{F:naive_example} using the realistic and the simplified simulation. The results are also compared with the simulation of the circuit without the QHE scheme.}
	\label{F:example_results}
\end{figure}

The results are shown in Figure \ref{F:example_results}. Both QHE simulations provide the homogeneous distribution before decrypting, and the same results as the original unencrypted circuit after decrypting.

\section{Update circuit for the bipartite graph}

\begin{figure}[H]
	\centering
	\subfigure[]{\includegraphics[scale=0.55]{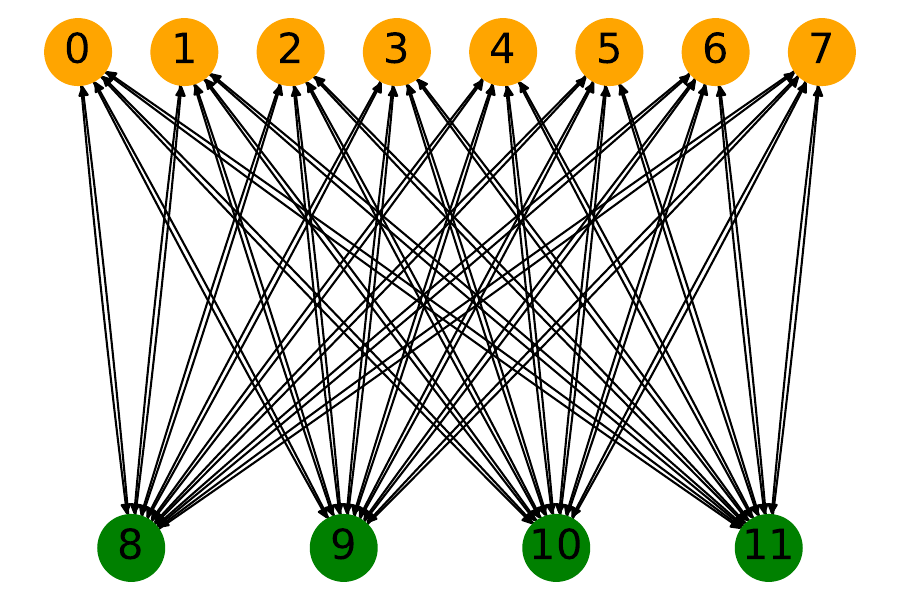}\label{F:bipartite_graph_SM}}
	\subfigure[]{\includegraphics[scale=0.55]{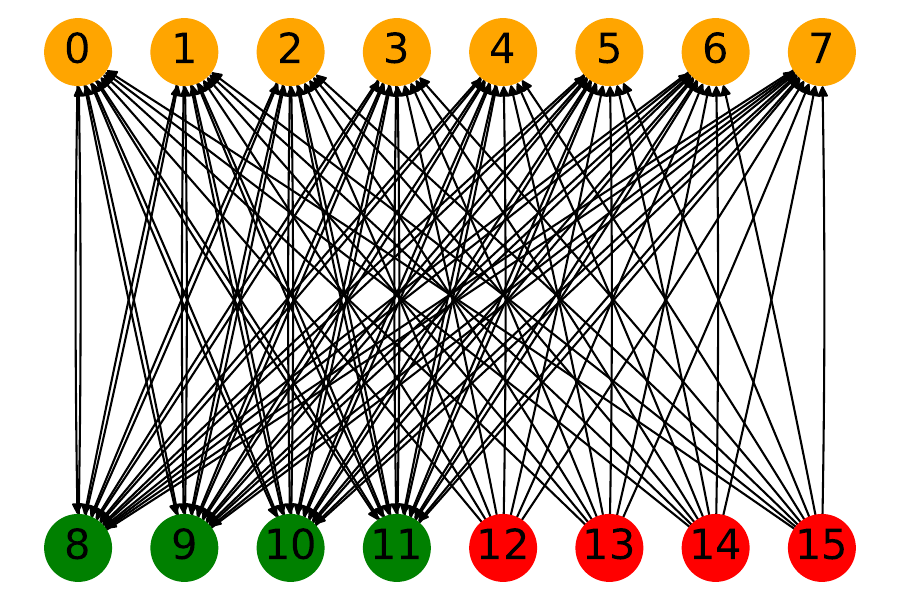}\label{F:bipartite_graph_augmented}}
	\caption{a) Representation of a complete bipartite graph with a set of $N_1=8$ nodes shown in orange and another set of $N_2=4$ nodes shown in green. b) Representation of an augmented complete bipartite graph with a third set of $N_1-N_2=4$ nodes shown in red.}
	\label{F:bipartite_graphs_SM}
\end{figure}

A complete bipartite graph is shown again in Figure {\color{blue} S}\ref{F:bipartite_graph_SM}. It is formed by two sets of nodes with $N_1$ and $N_2$ nodes, so that it has $N = N_1 + N_2$ nodes. Each node links to all the nodes of the other set, but there are no edges connecting two nodes inside the same set. Let $n_1$ and $n_2$ be two integer numbers so that $N_1 = 2^{n_1}$ and $N_2 = 2^{n_2}$, with $n_2 \leq n_1$. In order to construct a quantum circuit for this graph we need $n = n_1+1$ qubits, so that the circuit simulates Szegedy quantum walk in an augmented graph with $2^n = 2N_1 \geq N_1+N_2$ nodes, which is shown in Figure {\color{blue} S}\ref{F:bipartite_graph_augmented}. This graph has a third set with $N_1 - N_2$ additional nodes shown in red, which link to the $N_1$ nodes of the first set. However, no node in the graph links to them.

The Hilbert space of the original bipartite graph with $N_1+N_2$ nodes is
\begin{equation}
\mathcal{H} = \text{span}\lbrace\left|i\right>_1\left|j\right>_2,\ i,j = 0,1,...,N_1+N_2-1\rbrace.
\end{equation}
We will prove that it can be recovered as an invariant subspace of the Hilbert space of the augmented graph, which is
\begin{equation}
\widetilde{\mathcal{H}} = \text{span}\lbrace\left|i\right>_1\left|j\right>_2,\ i,j = 0,1,...,2N_1-1\rbrace.
\end{equation}

Let us denote the set of indexes for the first set of $N_1$ nodes as $\vertexset_1$, for the second set of $N_2$ nodes as $\vertexset_2$, and for the new third set of $N_1-N_2$ nodes as $\vertexset_3$. The transition matrix for the augmented graph is
\begin{equation}\label{G_bipartite_augmented}
\widetilde{G}_{ji} =
\left\lbrace\begin{array}{c}
\displaystyle \ \ \ \ \frac{1}{N_2} \ \ \ \ \ \ \text{if} \ i \in \vertexset_1 \ \text{and} \ j \in \vertexset_2,\\
\\
\displaystyle \ \ \ \ \frac{1}{N_1} \ \ \ \ \ \ \text{if} \ i \in \vertexset_2 \ \text{and} \ j \in \vertexset_1,\\
\\
\displaystyle \ \ \ \ \frac{1}{N_1} \ \ \ \ \ \ \text{if} \ i \in \vertexset_3 \ \text{and} \ j \in \vertexset_1,\\
\\
\displaystyle 0 \ \ \ \ \ \ \ \ \ \ \ \ \ \ \  \text{otherwise}.
\end{array}
\right.
\end{equation}
The unitary evolution operator for Szegedy quantum walk in this graph is
\begin{equation}\label{U_bipartite_augmented}
\widetilde{U}_w = S_w\left[2\sum_{i=0}^{2N_1-1} \left|\psi_i\right>\left<\psi_i\right| - \mathbbm{1}\right] = S_w\left[2\sum_{i \in (\vertexset_1 \cup \vertexset_2)} \left|\psi_i\right>\left<\psi_i\right| + 2\sum_{i \in \vertexset_3} \left|\psi_i\right>\left<\psi_i\right| - \mathbbm{1}\right],
\end{equation}
where we have separated the sum in a term which corresponds to the $\left|\psi_i\right>$ states of the original nodes, and another term with the $\left|\psi_i\right>$ of the added nodes. From equation (27) and \eqref{G_bipartite_augmented} we have that
\begin{equation}\label{psi_i_SM}
\left|\psi_i\right> =
\left\lbrace\begin{array}{c}
\displaystyle \left|i\right>_1 \otimes \sum_{k \in \vertexset_2}\frac{1}{\sqrt{N_2}}\left|k\right>_2 \ \ \ \ \ \ \text{if} \ i \in \vertexset_1,\\
\\
\displaystyle \left|i\right>_1 \otimes \sum_{k \in \vertexset_1}\frac{1}{\sqrt{N_1}}\left|k\right>_2 \ \ \ \ \ \ \text{if} \ i \in \vertexset_2,\\
\\
\displaystyle \left|i\right>_1 \otimes \sum_{k \in \vertexset_1}\frac{1}{\sqrt{N_1}}\left|k\right>_2 \ \ \ \ \ \ \text{if} \ i \in \vertexset_3.
\end{array}
\right.
\end{equation}
If we take an arbitrary state in the original Hilbert space $\mathcal{H}$, it is a linear combination of the computational basis whose first register has an index in $\vertexset_1 \cup \vertexset_2$. Thus, it is perpendicular to the $\left|\psi_i\right>$ states for $i \in \vertexset_3$. Therefore, the second sum term in \eqref{U_bipartite_augmented} does not intervene and the action is the same as that of the original graph evolution operator:
\begin{equation}
U_w = S_w\left[2\sum_{i=0}^{N_1+N_2-1} \left|\psi_i\right>\left<\psi_i\right| - \mathbbm{1}\right].
\end{equation}
Thus, the original Hilbert space $\mathcal{H}$ is invariant under the action of $\widetilde{U}_w$, and the action is the same as the original quantum walk. This is so because the extra $N_1-N_2$ nodes are not linked by any node. However, it does not matter what nodes they link to. It is chosen this way for convenience in order to construct a quantum circuit for the update operator.\\

\begin{figure}[H]
	\centering
	\includegraphics[scale=1]{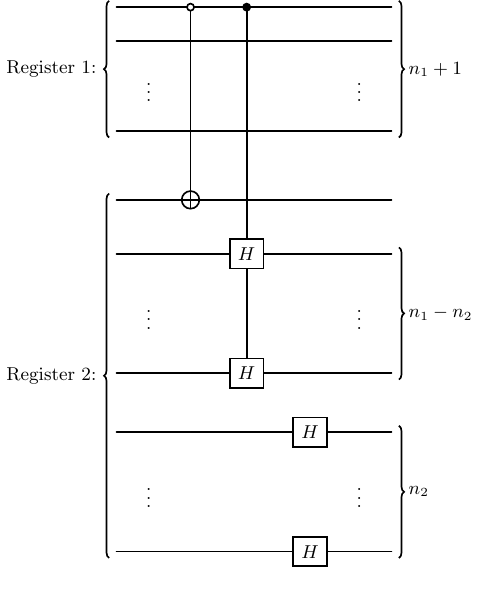}
	\caption{Quantum circuit for the update operator $\updateSM$ for Szegedy quantum walk over a complete bipartite graph.}
	\label{F:update_bipartite_SM}
\end{figure}

The action of the update operator $\updateSM$ is defined as
\begin{equation}
\updateSM\left|i\right>_1\left|0\right>_2 = \left|\psi_i\right>.
\end{equation}
The quantum circuit that we propose for implementing it is shown in Figure \ref{F:update_bipartite_SM}. For an initial state $\left|i\right>_1\left|0\right>_2$ with $i \in \vertexset_1$, the first qubit of the first register is in state $\left|0\right>$. Thus, an $X$ gate is applied to the first qubit of the second register, letting it in $\left|1\right>$. Any computational basis state with the first qubit in $\left|1\right>$ corresponds to the second half of nodes. With the Hadamard gates applied over the last $n_2$ qubits, the second register ends up in a linear combination of the first $2^{n_2}=N_2$ states $\left|k\right>_2$ of the second half of nodes. Thus, it is a linear combination of all the nodes in $\vertexset_2$ and the $\left|\psi_i\right>$ state for $i \in \vertexset_1$ is obtained. If the initial state is $\left|i\right>_1\left|0\right>_2$ with $i \in \vertexset_2$ or $i \in \vertexset_3$, the first qubit of the first register is in state $\left|1\right>$. In this case, Hadamard gates are applied to the last $n_1$ qubits of the second register, so it ends up in a linear combination of $2^{n_1}=N_1$ states whose first qubit is in state $\left|0\right>$. Therefore, it corresponds to the linear combination of all the states associated to the nodes in $\vertexset_1$, and again the corresponding $\left|\psi_i\right>$ are obtained. Then, this quantum circuit effectively implements the action of the update operator $\updateSM$.

\end{document}